\documentclass[twoside]{article}
\usepackage[utf8]{inputenc}
\usepackage{amsmath}
\usepackage{amssymb}
\usepackage{amsthm}
\usepackage[square,sort,numbers]{natbib}
\usepackage{xparse}
\usepackage{physics}
\usepackage{mathtools}
\usepackage{easybmat}
\usepackage{bm}
\usepackage{enumitem}
\usepackage{changepage}

\usepackage{mwe}
\usepackage{multicol}
\usepackage{float}
\usepackage{subcaption}
\usepackage[font=normalsize,labelfont=bf]{caption}
\usepackage{listings}
\usepackage{breqn}
\usepackage{braket}
\usepackage{graphicx, wrapfig}
\usepackage{xcolor}
\usepackage[section]{placeins}
\lstdefinestyle{base}{
	language=C,
	emptylines=1,
	breaklines=true,
	basicstyle=\ttfamily\color{blue},
	moredelim=**[is][\color{red}]{@}{@},
}

\usepackage{hyperref}

\hypersetup{
    colorlinks,
    pagebackref,
    pdfusetitle,
    urlcolor=darkblue,
    citecolor=darkblue,
    linkcolor=darkred,
    bookmarksnumbered,
    plainpages=false
}

\usepackage{caption}
\usepackage{amsmath,amsfonts,graphicx}
\usepackage{amssymb}
\usepackage{listings}
\usepackage{algorithmic}
\usepackage{scrextend}
\usepackage{color}   

\newcounter{lecnum}
\renewcommand{\thepage}{\thelecnum-\arabic{page}}

\definecolor{OliveGreen}{cmyk}{0.64,0,0.95,0.40}
\definecolor{CadetBlue}{cmyk}{0.62,0.57,0.23,0}
\definecolor{lightlightgray}{gray}{0.93}
\definecolor{darkblue}{rgb}{0,0,.6}
\definecolor{darkred}{rgb}{.7,0,0}
\definecolor{darkgreen}{rgb}{0,.6,0}
\definecolor{red}{rgb}{.98,0,0}

\newcommand{\lecture}[8]{
    \pagestyle{myheadings}
    \thispagestyle{plain}
    \newpage
    \setcounter{lecnum}{#1}
    \setcounter{page}{1}
    \setcounter{section}{0}
    \setcounter{figure}{0}
    \noindent
    \begin{center}
    \framebox{
        \vbox{\vspace{2mm}
        \hbox to 6.28in { {\bf Quantum Integer Programming 
    		\hfill September-December 2020} }
        \vspace{4mm}
        \hbox to 6.28in { {\Large \hfill Lecture #1: #2  \hfill} }
        \vspace{2mm}
        \hbox to 6.28in { {\it Lecturer: #3 \hfill Scribe: #4} }
        \vspace{2mm}
        \hbox to 6.28in { \href{ #5 }{Slides}, \href{ #6 }{Notebook}   \hfill \href{ #7 }{Video}, password: \text{#8} }
        \vspace{2mm}}
       }
    \end{center}
    \markboth{Lecture #1: #2}{Lecture #1: #2}
}

\theoremstyle{plain}
\newtheorem{theorem}{Theorem}

\theoremstyle{definition}
\newtheorem{definition}{Definition}

\newtheorem{example}{Example}

\theoremstyle{remark}

\newtheorem*{note}{Note}

\begin{document} 
\begin{center}
\section*{Quantum Integer Programming (QuIP) 47-779: Lecture Notes\\
David E. Bernal, Sridhar Tayur, Davide Venturelli \\
Fall 2020
}
\subsection*{Abstract}
\end{center}
This lecture series on Quantum Integer Programming (QuIP) -- created by Professor Sridhar Tayur, David E. Bernal and Dr. Davide Venturelli, a collaboration between CMU and USRA, with the support from Amazon Braket during Fall 2020 -- is intended for students and researchers interested in Integer Programming and the potential of near term quantum and quantum inspired computing in solving optimization problems.

Originally created for Tepper School of Business course 47-779 (at CMU), these were also used for the course ID5840 (at IIT-Madras, by Professors Anil Prabhakar and Prabha Mandayam) whose students (listed at the beginning of each lecture) were scribes. Dr. Vikesh Siddhu, post-doc in CMU Quantum Computing Group, assisted during the lectures, student projects and with proof-reading this scribe.

Through these lectures one will learn to formulate a problem and map it to a Quadratic Unconstrained Binary Optimization (QUBO) problem, understand various mapping and techniques like the Ising model, Graver Augmented Multiseed Algorithm (GAMA), Simulated or Quantum Annealing and QAOA, and ideas on how to solve these Integer problems using these quantum and classical methods.

The course website (with lecture videos and colab notebooks):  
\url{https://bernalde.github.io/QuIP/}

\textbf{Keywords:} Ising model, Integer Programming, Computational Algebraic Geometry, Graver Basis, Quantum Annealing, Simulated Annealing, Combinatorial Optimization, Graph coloring, discrete nonlinear optimization.

\setitemize{noitemsep,topsep=0pt,parsep=0pt,partopsep=0pt}
\begin{center}
{\Large \textsc{47-779. Quantum Integer Programming}}
\end{center}
\begin{center}
Mini-1, Fall 2020
\end{center}

\begin{center}
\rule{6in}{0.4pt}
\begin{minipage}[t]{.75\textwidth}
\begin{tabular}{llr}
\textbf{Room:} Zoom Online & & \textbf{Time:} Tuesday and Thursday 5:20pm-7:10pm
\end{tabular}
\end{minipage}
\rule{6in}{0.4pt}
\end{center}
\setlength{\unitlength}{1in}
\renewcommand{\arraystretch}{2}

\noindent\textbf{Instructors:}
\begin{center}
\rule{6in}{0.4pt}
\begin{minipage}[t]{.85\textwidth}
\begin{tabular}{lll}
Sridhar Tayur & \textbf{Email:}  \href{mailto:stayur@cmu.edu}{stayur@cmu.edu} & \textbf{Office:} 4216 Tepper Quad \\
\hline
Davide Venturelli & \textbf{Email:}  \href{mailto:DVenturelli@usra.edu}{DVenturelli@usra.edu} & \textbf{Office:} Online \\
\hline
David E. Bernal & \textbf{Email:}  \href{mailto:bernalde@cmu.edu}{bernalde@cmu.edu} & \textbf{Office:} 3116 Doherty Hall
\end{tabular}
\end{minipage}
\rule{6in}{0.4pt}
\end{center}

\noindent\textbf{Course Webpage:}
\begin{itemize}
\item \url{https://bernalde.github.io/QuIP/}
\end{itemize}


\vskip.15in
\noindent\textbf{Objectives:}  This course is primarily designed for graduate students (and advanced undergraduates) interested in integer programming (with non-linear objective functions) and the potential of near-term quantum and quantum-inspired computing for solving combinatorial optimization problems.
By the end of the semester, someone enrolled in this course should be able to:
\begin{itemize}
    \item Appreciate the current status of quantum computing and its potential use for integer programming
    \item Access and use quantum computing resources (such as D-Wave Quantum Annealers)
    \item Set up a given integer program to be solved with quantum computing 
    \item Work in groups collaboratively on a state-of-the-art project involving applications of quantum computing and integer programming
\end{itemize}

\noindent This course is not going to focus on the following topics:
\begin{itemize}
    \item Quantum Gates and Circuits
    \item Computational complexity theory
    \item Quantum Information Theory
    \item Analysis of speedup using differential geometry, algebraic topology, etc.
\end{itemize}

\vskip.15in
\noindent\textbf{Prerequisite classes and capabilities:}
Although this class has no explicit prerequisites we consider a list of recommended topics and skills that the student should feel comfortable with.
An undergraduate-level understanding of probability, calculus, statistics, graph theory, algorithms, and linear algebra is assumed.
Knowledge of linear and integer programming will be useful for this course.
Programming skills are strongly recommended.
Basic concepts in physics are recommended but lack of prior knowledge is not an issue as pertinent ones will be covered in the lectures.
No particular knowledge in quantum mechanics or algebraic geometry is required. 

\vspace*{.15in}

Students with backgrounds in operations research, industrial engineering, chemical engineering, electrical engineering, physics, computer science, or applied mathematics are strongly encouraged to consider taking this course.


\noindent \textbf{Tentative Course Outline:}
\begin{center} 
\begin{minipage}{5in}
\begin{flushleft}
{\bf Part 1 - Integer programming (classical methods)} \dotfill ~1 week \\
\begin{itemize}
\item Integer Programming basics \cite{conforti2014integer}. \\
\item Cutting plane theory and relaxations \cite{conforti2014integer}. \\
\item Introduction to Test Sets  \cite{weismantel1998test,sturmfels1996grobner,tayur1995algebraic,de2012algebraic}. \\
\begin{itemize}
    \item Gr{\"o}bner basis \cite{bertsimas2000new,conti1991buchberger,hocsten1995grin}. \\
    \item Graver basis \cite{hemmecke2011polynomial}. \\
\end{itemize}
\end{itemize}
{\bf Part 2 - Ising, QUBO} \dotfill ~1 week \\

\begin{itemize}
\item Ising model basics \cite{brush1967history,sherrington1975solvable,ray1989sherrington}. \\
\item Simulated Annealing \cite{kirkpatrick1983optimization,koulamas1994survey}. \\
\item Markov-chain Monte Carlo methods \cite{metropolis1953equation,bortz1975new,troyer2005computational,young2008size}. \\
\item Benchmarking classical methods \cite{dunning2018works,coffrin2019evaluating}. \\
\item Formulating combinatorial problems as QUBOs \cite{lucas2014ising}. \\

\end{itemize}

{\bf Part 3 - GAMA: Graver Augmented Multiseed algorithm} \dotfill ~1 week \\
\begin{itemize}
\item GAMA \cite{alghassi2019graver}. \\
\begin{itemize}
\item Applications: Portfolio Optimization \cite{alghassi2019graver}, Cancer Genomics \cite{alghassi2019quantum} \\
\item Quantum Inspired: Quadratic (Semi-)Assignment Problem \cite{alghassi2019gama}.
\end{itemize}
\end{itemize}

{\bf Part 4 - Quantum methods for solving Ising/QUBO} \dotfill ~1 week \\
\begin{itemize}
\item AQC, Quantum Annealing and D-Wave \cite{mcgeoch2020theory,albash2018adiabatic,das2008colloquium,santoro2006optimization,farhi2001quantum,kadowaki1998quantum,johnson_quantum_2011}. \\
\item QAOA: Quantum Approximate Optimization Algorithm \cite{farhi2014quantum,hadfield2017quantum,hadfield2019quantum}. \\

\end{itemize}

{\bf Part 5 - Hardware for solving Ising/QUBO} \dotfill ~1 week \\
\begin{itemize}
\item Graphical Processing Units \cite{yavorsky2019highly,cook2019gpu,romero2019performance}. \\
\item Tensor Processing Units \cite{yang2019high}. \\
 \item Complementary metal-oxide-semiconductors (CMOS) \cite{yamaoka201520k}. \\
\item Digital Annealers \cite{aramon2019physics}. \\
\item Oscillator Based Computing \cite{chou2019analog,wang2019oim}. \\
\item Coherent Ising Machines \cite{roques2020heuristic,inagaki2016coherent,king2018emulating,hamerly2019experimental,tiunov2019annealing,mcmahon2016fully}. \\

\end{itemize}

{\bf Part 6 - Other topics and project presentations} \dotfill ~1 week \\
\begin{itemize}
\item Compiling \\

\begin{itemize}
\item Quantum Annealing \cite{bernal2019integer,dridi2018novel}. \\
\item Gate-based  Noisy Intermediate Scale Quantum (NISQ) devices \cite{dridi2019knuth}. \\
\end{itemize}
\item Adiabatic Quantum Computing and Algebraic Geometry \\
\begin{itemize}
\item Minimizing Polynomial Functions \cite{dridi2019minimizing}.\\
\item Prime Factorization \cite{dridi2017prime}.
\end{itemize}

\end{itemize}
\end{flushleft}
\end{minipage}
\end{center}


\vspace*{.15in}
\noindent\textbf{Project description:} 
This project will be completed in groups of 2-4 students and will reflect the understanding of the students of the material covered in the lecture. The components of this project are the following
\begin{itemize}
    \item Identify a problem that can be posed as an Integer Program. Discuss the importance of this problem.
    \item Solve instances of the identifies problem using classical tools. Identify which are the sources of complexity while solving this problem.
    \item Model the problem as a Quadratic Unconstrained Binary Optimization (QUBO). Verify that the reformulation of the problem is valid, in the sense that it represents the original problem.
    \item Solve the resulting QUBO using non-conventional methods, e.g. Quantum Annealing, QAOA, simulated annealing in GPUs/TPUs, etc. Compare at least two different methods.
    \item Write a report outlining the different approaches used and highlighting the knowledge obtained while developing the project.
    \item Hold a final presentation in front of the class reporting the findings of the project.
\end{itemize}



\vskip.15in
\noindent\textbf{Highlights:}
The specific skills that students will gain that will make them ``quantum ready'' for industry or further academic research in this course are:
\begin{enumerate}
    \item Classical
    \begin{enumerate}
        \item Given a practical problem (from supply chain or physics or anything else), formulate it as a non-linear integer program. We will provide a few practical problems, but we encourage you to suggest one that you are already working on or are interested in.
        \item Solve such formulations via classical solvers.
    \end{enumerate}
    \item Quantum
    \begin{enumerate}
        \item Reformulate the problem to be “quantum ready” by making it in the form of a QUBO.
        \item Solve the QUBO “brute force” through D-Wave or IBM (via QAOA).
    \end{enumerate}
    \item Hybrid Quantum-Classical
    \begin{enumerate}
        \item Reformulate the problem again in the form suitable for GAMA.
        \item Solve GAMA compatible formulation via D-Wave and/or via QAOA.
    \end{enumerate}
\end{enumerate}
{\bf USRA Collaboration}
\begin{enumerate}
    \item Access to D-Wave systems might be available via written proposals to the University Space Research Association (USRA). See \url{https://riacs.usra.edu/quantum/rfp} for terms and conditions. The course will discuss proposal preparation.
    \item Students of this course are encouraged to apply to the Feynman Academy Internship program \url{https://riacs.usra.edu/quantum/qacademy} that sponsors research projects at NASA Ames Research Center.
\end{enumerate}
{\bf Amazon Braket Collaboration}
\begin{enumerate}
    \item Access to D-Wave, Rigetti and ionQ have been made available by Amazon Braket.
    \item Students will be provided individual sub-accounts for use with pre-funded amounts to access the quantum machines.
\end{enumerate}


\vskip.15in
\noindent\textbf{Casual References:} 
There is no single text book for the course. This is a short list of various interesting and useful books that will be mentioned during the course. You need to consult them occasionally.
\begin{itemize}
\item Georges Irfah, {\textit {The Universal History of Computing}}, John Wiley \& Sons, 2001.
\item A. Das and B.K. Chakrabarti (Eds.). {\textit {Quantum Annealing and Related Optimization Methods}, Springer-Verlag, 2005}.
\item Eleanor G. Rieffel and Wolfgang H. Polak, {\textit{Quantum Computing: A Gentle Introduction}}, MIT Press, 2011.
\item Richard J. Lipton and Kenneth W. Regan, {\textit{Quantum Algorithms via Linear Algebra. A Primer}}, MIT Press, 2014.
\end{itemize} 




\lecture{1}
{Integer Programming}
{Prof. Sridhar Tayur, David E. Bernal}
{Vighnesh Natarajan}
{https://bernalde.github.io/QuIP/slides/47-779\%20Lecture\%201\%20-\%20Integer\%20Programming.pdf}
{https://colab.research.google.com/github/bernalde/QuIP/blob/master/notebooks/Notebook\%201\%20-\%20LP\%20and\%20IP.ipynb}
{https://cmu.zoom.us/rec/play/aYR6uCs3R0vdyfdYJ5uiJpnyUgb5CuLB_Jebl37M_rfKE5EMvNl5u7TDXadz8RmST45JrSXBBN1UYYMl.LnkHOy7ZB34jirzn}
{P^2GSWny}

\section{Introduction}
The standard approach to solving real world optimizations
is as follows:
\begin{enumerate}
\item Mathematically model the problem
\begin{itemize}
      \item Quantify the real world objective as a mathematical function
      \item Locate variables on which this function depends 
      \item List real world constraints as constraints
      on the variables
  \end{itemize}
  \item Solve the model, i.e., find the best objective value and variables
  \item Interpret Results
\end{enumerate}
Using the first step above, most real world problems are brought to a mathematical model of the form:
\begin{align*}
    \underset{x}{\text{minimize}} \;& f(x)\\
    \text{subject to: } & g(x) \leq 0\\
               & h(x) = 0
\end{align*}
In this model, we call $f(x)$ the objective function, $g(x) \leq 0$ an inequality constraint, and $h(x) = 0$ an equality constraint.
Such models are called often called {\em programs}. These programs are categorized based on the objective function and constraints, and the nature of the variables. Some popular categories include Linear Programs, Convex Programs, and Integer Programs. We shall discuss such
programs in more depth, especially Integer Programs.

\section{Linear Programming}
Suppose there is a company that produces two different products, A and B, which can be sold at different values, \$5.5 and \$2.1 per unit, respectively. The company only counts with a single machine with electricity usage of at most 17kW/day, and producing each A and B consumes 8kW/day and 2kW/day, respectively. Besides, the company can only produce at most 2 more units of A than B per day.
This real world problem can be mathematically modelled as follows.

Assuming the units produced of A are $x_1$ and of B are $x_2$ we have
\begin{align*}
    \max_{x_1, x_2}& 5.5x_1 + 2.1x_2 \\
    s.t. & x_2 \leq x_1 + 2 \\
    &8x_1 + 2x_2 \leq 17 \\
    &x_1, x_2 \geq 0
\end{align*}

 \begin{figure}[htb]
 \centering
 \includegraphics[scale=0.5]{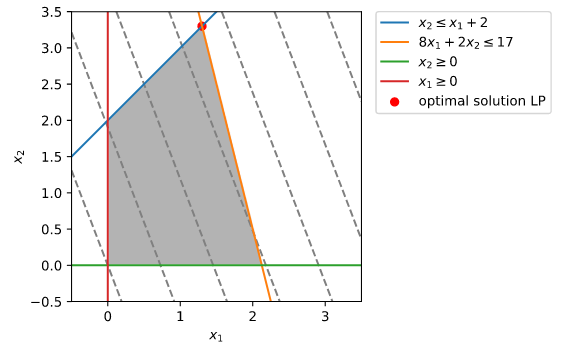}   
 \caption{Feasible region and solution of sample LP}
 \label{fig:LP}
 \end{figure}

The model mentioned above is an example of a linear program~(LP). In an LP, the variables are real, continuous, and satisfy affine constraints. The objective function depends linearly on these variables. As a result, the general form of an LP is
\begin{align*}
    \min_{x}&\; & c^T x\\
    subject\: to: \;& Ax \leq b\\
\end{align*}
In an LP, the feasible region, i.e.,  the set variables which satisfy the problem's constraints, is a convex polyhedron. Such
polyhedron are an intersection of half-spaces and hyperplanes.

The standard ways to solve an LP include
\begin{itemize}
    \item Simplex methods
    \item Interior point methods
\end{itemize}

\subsection{The simplex method}

This is a vertex hopping method. The method provides a worst-case exponential time complexity. A $k$-simplex is, in essence, a convex hull of $k+1$ points. The optimal solution occurs at the vertices of the simplex, where some of the constraints are active. Thus we find the optima by vertex hopping. The standard LP can be expressed as a matrix, which is then used to solve the problem

\[ 
\left( \begin{array}{cccc}
1 & -c^T & 0 \\
0 & A & b
\end{array} \right)
\longrightarrow
\left( \begin{array}{cccc}
1 & -c_B^T & -c_N^T & 0 \\
0 & I & D & b
\end{array} \right)
\longrightarrow
\left( \begin{array}{cccc}
1 & 0 & -c_N^T & z_b \\
0 & I & D & b
\end{array} \right)
\]
Here, we refer to the basic variables, $-c_N^T$ is the relative cost, $z_b$ is the objective value, the subscript $B$ denotes the basic variables values, while $D$ denotes the non-basic variables. The simplex method is implemented in a variety of solvers including Gurobi\footnote{https://www.gurobi.com/}, IBM CPLEX\footnote{https://www.ibm.com/analytics/cplex-optimizer}, XPRESS\footnote{https://www.fico.com/en/products/fico-xpress-solver}, MindOpt\footnote{https://solver.damo.alibaba.com/htmlpages}, CBC\footnote{https://github.com/coin-or/Cbc}, GLPK\footnote{https://www.gnu.org/software/glpk/}, among others.

\subsection{Interior point methods}
These methods start with a point inside the feasible region and iteratively taking steps to never leave the feasible region. This contrasts with optimization algorithms like projected gradient descent, projected steepest descent, primal-dual optimization, etc.; where a step can leave the feasible set, and later be projected back onto the feasible set.
We start from a feasible solution and get closer and closer to the edges by having a monotonic barrier when close. Thus we never reach the edge, just get closer and closer to activating the appropriate constraint. This method is also known as a barrier method and is very useful when problems are dual degenerate.

Examples of interior-point solvers include  Gurobi\footnote{https://www.gurobi.com/}, IBM CPLEX\footnote{https://www.ibm.com/analytics/cplex-optimizer}, MOSEK\footnote{https://www.mosek.com/}, and COPT\_LP\footnote{https://www.shanshu.ai/copt/}, among others.

\section{Duality}
Every problem has a dual problem that can be expressed using a {\em Lagrangian}. The original problem, that has been described until now is known as the primal. For this primal problem:
\begin{align*}
    \underset{x}{\text{minimize}} \;& f(x)\\
    \text{subject\: to: }\; & g(x) \leq 0\\
               & h(x) = 0,
\end{align*}
define a Lagrangian,
\begin{align*}
    L(x, \lambda, \mu) = f(x) + \lambda^Tg(x) + \mu^T h(x)\\
    \lambda \geq 0.
\end{align*}
We define 
\begin{align*}
    \phi(\lambda, \mu) = \underset{x}{\text{argmin}}\: L(x, \lambda, \mu)
\end{align*}
This function bounds from below all feasible values of the primal objective $f$. The dual problem is to maximize $\phi(\lambda, \mu)$. However, a question arises how the maximum dual value $\phi^*$ is related to minimum value $f^*$ of the primal problem. There are two possibilities - 
\begin{itemize}
    \item Weak duality: $\phi^* \leq f^*$
    \item Strong duality: $\phi^* = f^*$
\end{itemize}
This can be proven using the Minimax theorem. However in cases of an LP, the Minimax theorem can be used to prove strong duality.

For convex optimization problems, where the objectives and inequality constraints in standard form have to be convex, and the equality constraints have to be affine, Slater’s conditions are a sufficient condition for strong duality to hold.\\
A short example has been illustrated below for an LP.
\begin{align*}
    \underset{x}{\text{minimize}}\; & c^T x\\
    \text{subject\: to: }\;& Ax \leq b\\
    & x \geq 0, x\:\epsilon\: \mathbb{R}^n
\end{align*}
The Lagrange function becomes $L(x, \lambda) = c^T x + \lambda^T(b-Ax), x \geq 0$
\begin{align*}
    \phi(\lambda) & = \underset{x\geq0}{\text{min }}  c^T x + \lambda^T(b-Ax)\\
    & = \lambda^T b + \underset{x\geq0}{\text{min }} (c^T - \lambda^T A)x\\
    \phi(\lambda) & = \lambda^T b, & \text{if}\; c^T - \lambda^T A \geq 0 & \\
    & = -\infty, & \text{otherwise} &\\
\end{align*}
Thus, our dual problem becomes maximizing $\phi(\lambda)$, which is $\lambda^T b$. And we have strong duality holding in LPs. So, 
\begin{align*}
    \lambda_{opt}^Tb = c^T x_{opt}
\end{align*}
With this, we can move on to non-linear optimizations.
\section{MIP and MINLP}
Re-visiting our sample problem on the company producing products, What if we have a constraint that lets us produce only integer number of products. This is something we need to modify and include in our formulation.
\begin{align*}
    \max_{x_1, x_2}\;& 5.5x_1 + 2.1x_2 \\
    s.t.\;& x_2 \leq x_1 + 2 \\
    &8x_1 + 2x_2 \leq 17 \\
    &x_1, x_2 \geq 0 \\
    &x_1, x_2\; \text{are integers}
\end{align*}
 \begin{figure}[htb]
 \centering
 \includegraphics[scale=0.5]{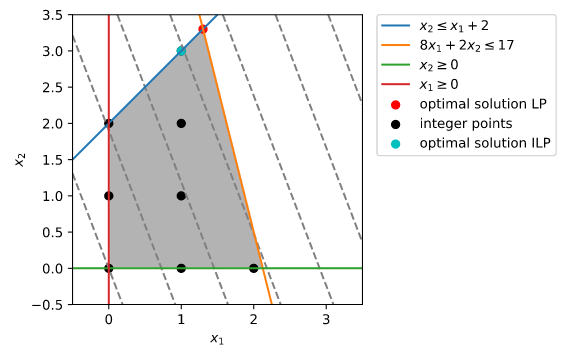}   
 \caption{Feasible region and solution of sample ILP}
 \label{fig:ILP}
 \end{figure}
\noindent The usual method to solve an Integer program is to first relax to a LP problem, and then round off to the nearest integer. However, there are better techniques that provide more accurate solutions.

Before proceeding, we need to understand the challenges and the need for good Integer solvers. How hard can it be to just enumerate all solutions? This depends on the complexity of the problem. Integer programs generally have exponential complexity. This complexity issue is clearly understood from the image below.
\begin{figure}[htb]
\centering
\includegraphics[scale=0.5]{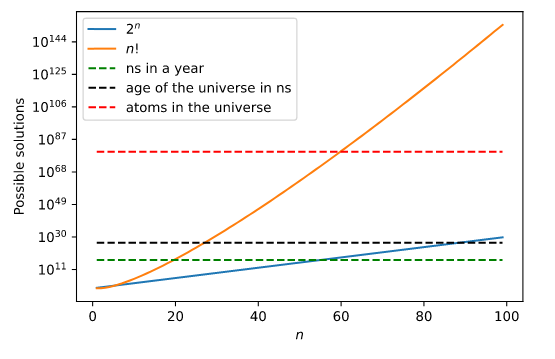}   
\caption{Enumeration of feasible solutions}
\label{fig:Enumerate}
\end{figure}
\noindent A Mixed-Integer Program (MIP) is an optimization problem of the form
\begin{align*}
    \underset{x}{minimize} \;& f(x)\\
    subject\: to:\; & g(x) \leq 0\\
               & \text{some or all }x_i's\text{ are integers}
\end{align*}
A MIP is a NP complete problem, and hence for large problem sizes, we do not have the means to solve it. Various techniques to find approximate solutions, and if possible exact solutions are
\begin{itemize}
    \item Branch and Bound - Solution of each search node using linear programming
    \item Cutting plane methods - based on polyhedral theory
\end{itemize}
In branch and bound, we essentially ask a set of ``yes - no'' questions. Answers to these questions help divide the feasibly region, and also help prune it for optimal solutions. There are various ways we can prune a set, by integrality, by bound and by infeasibility. The following image is an example of branch and bound on our sample MIP.
\begin{figure}[htb]
\centering
\includegraphics[scale=0.5]{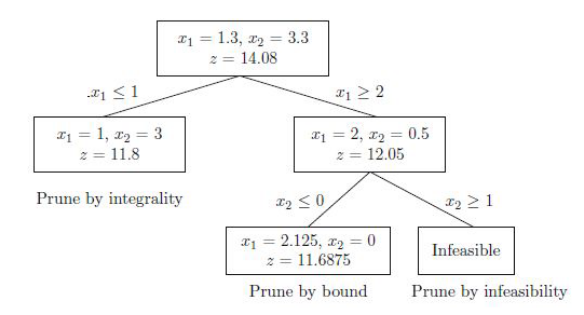}   
\caption{Branching and bounding on sample MIP \cite{conforti2014integer}}
\label{fig:bb}
\end{figure}
\noindent Cutting plane methods involve cutting off parts of the feasible region of the LP relaxed problem. This can be used for convex optimization problems also, with a convex boundary approximated by a set of linear boundaries. 

\noindent Recent improvements in MIP solvers are significant, with improvements in choosing cuts, pre-processing, branching variable selection, and heuristics. We see that CPLEX 11 (2007) is 29000 times faster than CPLEX 1.2 (1991). Gurobi 1 is comparable to CPLEX 11, and Gurobi 8 (2018) is 91 times faster. Overall we see roughly a total speedup of 2.6M times from 1991 to 2018. In addition, we have had huge growth in hardware speeds, which have also led to speedups. These solvers, which are now state of the art, are now exploring using specialized hardware such as parallel processing GPU techniques to push performance.

\noindent The real world is nonlinear. Most systems are linearized about operating points, from where we can apply linear programming and integer programming. However, there are cases where this linearization does not yield satisfactory answers, and we need to involve nonlinear characteristics. Thus our objective function is no more of the form $c^T x$, and our objective is no more of the form $Ax = b$. 

\noindent This brings us into the realm of Mixed Integer Non-Linear Programming or MINLP. MINLP are NP-hard problems. We can see that all the problems we have considered are all subsets of MINLP, which is the general and much more challenging set of problems. Similar to MIP, branching, and bounding is a method used to solve MINLP problems. Also, MIP based methods are used, with the nonlinear objectives linearized in two ways. The first way is where the objectives are under-estimated. The second way is where the objectives are overestimated. 

\begin{figure}[htb]
\centering
\begin{subfigure}{.5\textwidth}
  \centering
  \includegraphics[scale=0.5]{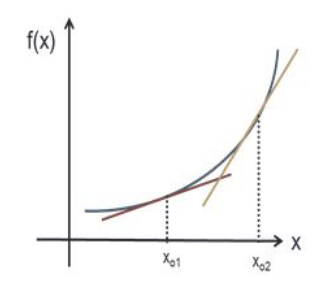}
  \caption{Under-estimate the cost function}
  \label{fig:under}
\end{subfigure}%
\begin{subfigure}{.5\textwidth}
  \centering
  \includegraphics[scale=0.5]{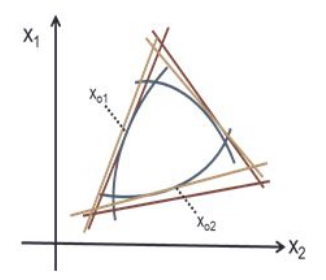}
  \caption{Over-estimate feasible region}
  \label{fig:over}
\end{subfigure}
\caption{Linearization of MINLP}
\label{fig:estimate}
\end{figure}

\noindent Continuing with our example of the two products, if we add an extra constraint, which is non linear as shown below, we can see how the feasible region changes. 
\begin{align*}
    \max_{x_1, x_2}\;& 5.5x_1 + 2.1x_2 \\
    s.t.\;& x_2 \leq x_1 + 2 \\
    &8x_1 + 2x_2 \leq 17 \\
    &(x_2-1)^2 \leq 2-x_1 \\
    &x_1, x_2 \geq 0 \\
    &x_1, x_2\text{ are Integers} 
\end{align*}
\begin{figure}[htb]
\centering
\includegraphics[scale=0.4]{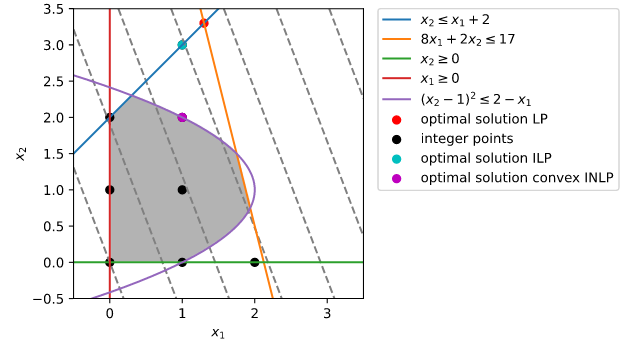}   
\caption{Feasible region and optimal solution of the convex MINLP}
\label{fig:cvxminlp}
\end{figure}

\noindent It is easy to observe that this particular non-linearity preserves convexity. Hence this is still a straightforward problem to solve. Things start to get tricky when the problem becomes non-convex. We shall see non-convexity in the next example. Solvers can still solve such non-convex problems. However, we can understand what can go wrong and where solvers can get stuck in these non-convex situations

We add an extra constraint to the formulation, as follows.
\begin{align*}
    \max_{x_1, x_2}\;& 5.5x_1 + 2.1x_2 \\
    s.t.\;& x_2 \leq x_1 + 2 \\
    &8x_1 + 2x_2 \leq 17 \\
    &(x_2-1)^2 \leq 2-x_1 \\
    &(x_2-1)^2 \geq 0.5+x_1 \\
    &x_1, x_2 \geq 0 \\
    &x_1, x_2\text{ are Integers} 
\end{align*}
\begin{figure}[htb]
\centering
\includegraphics[scale=0.4]{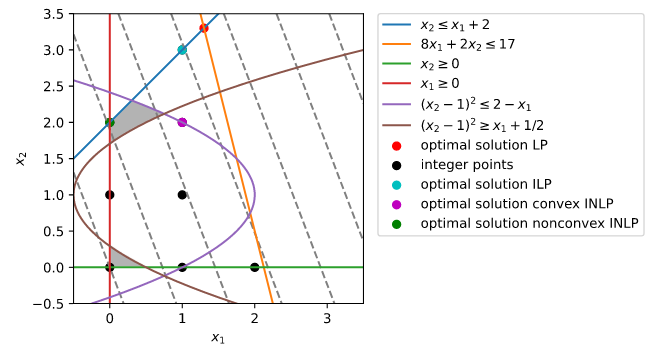}   
\caption{Feasible region and optimal solution of the non-convex MINLP}
\label{fig:non_cvxminlp}
\end{figure}

\noindent A state of the art nonconvex MINLP solver is Baron\footnote{https://minlp.com/baron}. This solver is benchmarked using MINLPLib with instances with approximately 2500 variables and constraints. MIP solvers like Gurobi, IBM-CPLEX, and XPRESS are increasing their capabilities to solve nonlinear problems also.

\section{Complexity Analysis}
Analysis of problems involves studying the time and memory required to solve them. The main focus here is the time complexity involved in solving problems. Problems are usually categorized by the worst-case scenario performance; however, some algorithms have become very popular due to their exceptional average-case performance, such as quick-sort. A standard notation involved here is the “Big - $O$” notation.

The "Big-$O$" notation is used where we have two functions $f: S \xrightarrow{} \mathbb{R}_+$ and $g: S \xrightarrow{} \mathbb{R}_+$, there exist positive numbers M and $x_o$, such that $f(x) \leq M g(x)\; \forall x \geq x_o$ we say $f(x) = O(g(x))$.\\\\
We say that $f(x)$ is polynomial time if it is polynomially bounded by $g(x)$, in order of complexity. For example, selection sort, is $f(n) = O(n^2)$, however for quick sort, $f(n) = O(n \log(n))$. If there exists a polynomial time solution for a problem, it belongs to the class P. LP problems are in the class P(Polynomial), because interior point algorithms solve LPs in polynomial time.

Decision problems are those that have only two possible answers, `Yes' or `No'. NP~(Non-deterministic Polynomial) class contains problems that cannot be solved in polynomial time complexity. However the problem solutions can be verified in polynomial time. That is, a `Yes' or `No' answer can be verified in polynomial time. A problem $Q$ can be called NP-hard if all problems in NP can be reduced to $Q$ in poly-time. Integer Programming is NP-hard. We can transform an NP problem into an integer program. Hence, if we find a way to solve IP in polynomial time, we can solve all NP problems in polynomial time and prove the result P=NP! Unfortunately, integer programs with quadratic constraints have been proven to be undecidable. Even after a long time without finding a solution, we cannot conclude that no solutions exists.

\noindent A problem is said to belong to the complexity class BPP (Bounded-error Probabilistic Polynomial time) if there is an algorithm that solves the problem in such a way that the algorithm 
\begin{itemize}
    \item is allowed to make random decisions.
    \item is guaranteed to run in polynomial time.
    \item given an incorrect answer with probability at most 1/3
\end{itemize}
\noindent BQP (Bound error Quantum Polynomial time) is a complexity class that is the quantum analog of BPP. The hope is that there are problems that belong to BQP and not in BPP, which will help in realizing quantum advantage~\citep{nielsen2002quantum}.


\lecture{2}
{Test-set methods - Gr{\"o}bner Basis (Part 1)}
{Prof. Sridhar Tayur, David E. Bernal}
{Srijan Gupta, Vighnesh Natarajan}
{https://bernalde.github.io/QuIP/slides/47-779\%20Lecture\%202\%20-\%20Test-set\%20methods.pdf}
{https://colab.research.google.com/github/bernalde/QuIP/blob/master/notebooks/Notebook\%202\%20-\%20Groebner\%20basis.ipynb}
{https://cmu.zoom.us/rec/play/Ihz6C8KxpIQ6Xnk_IXPEkJuso6D4eI_VDDGwT0tbGoqH9FyPC-Ip3B8mRLnUHSLfYO-zOZljg_DOX70Q.jaOZNfLdKHVEJqDk}
{6.gKTx4+}

In any interger program (IP), we have a set of constraints. This set is defined
using equality and inequality constraints. However, the inequality constraints
can be converted to equality constraints by adding new variables. Here, we
discuss a way to find all feasible solutions~(`points') to the set of
constraints.


The following section starts with the necessary preliminary mathematical
definitions: fun and deep in their respect. 

\section{Algebraic Geometry \& Polynomial Ring}

Algebraic geometry uses tools from algebra to study of geometric objects.
These geometric objects are defined by polynomial equations.

\subsection{Field, Ring, Affine-n space}

\begin{definition}{(Field)}
A \textbf{field} is a set $\mathbb{K}$ along with two operations defined on it:
`addition' ($a+b$) and `multiplication' ($a \cdot b$) for $a, b \in
\mathbb{K}$. Both these operations satisfy the usual field axioms:
associativity, commutativity, distributivity, existence of identity and
existence of inverse.

Examples of fields: the set of rational numbers $\mathbb{Q}$, and real numbers
$\mathbb{R}$.
\end{definition}

\begin{definition}{(Ring)}
A \textbf{ring} is a set $R$ which is closed under addition and multiplication.
This set forms an abelian group under addition, and it satisfies
associativity for multiplication and distributivity. 
\end{definition}

\begin{definition}{(Affine n-space)}
The \textbf{affine n-space} over the field $\mathbb{K}$, denoted by $\mathbb{A}^n_{\mathbb{K}}$ or simply $\mathbb{A}^n$, is the set of all n-tuples pf elements of $\mathbb{K}$. An element $P = (a_1, a_2,\dots,a_n) \in\mathbb{A}^n$ is called a \textbf{point}, and $a_i\in\mathbb{K}$ are called the \textbf{coordinates} of $P$.
\end{definition}

\subsection{Monomials, Polynomials, Polynomial Ring}

Let $\mathbb{K}$ be a field and $\vb{x} = x_1, x_2, \dots , x_n$ be indeterminates.

\begin{definition}{(Monomial)}
A \textbf{monomial} $\vb{x}^{\boldsymbol{\alpha}}$ is given by
        \begin{equation}
            \vb{x}^{\boldsymbol{\alpha}} \coloneqq  x_1^{\alpha_1}x_2^{\alpha_2}\cdots x_n^{\alpha_n}
        \end{equation}
where $\alpha_i$'s are non-negative integers. 
\end{definition}

\begin{definition}
The tuple $\boldsymbol{\alpha} = (\alpha_1, \alpha_2,\dots,\alpha_n)$ is called the \emph{multidegree} or \emph{exponent vector} of the monomial. The \textbf{degree of a monomial}, $\deg \boldsymbol{\alpha} = \sum_i \alpha_i$
\end{definition}

\begin{definition}{(Polynomial Ring)}
The \textbf{polynomial ring} $\mathbb{K}[\vb{x}]$ is (a ring of polynomials) the set of all \textbf{polynomials} $p(\vb{x})$ (finite linear combinations of the monomials $\vb{x}^{\boldsymbol{\alpha}}$) with coefficients in the field $\mathbb{K}$.
        \begin{equation}
            \mathbb{K}[\vb{x}] \coloneqq  \bigg\{p(\vb{x}) \text{ }\Big|\text{ }p(\vb{x}) = \sum_{\boldsymbol{\alpha}} c_{\boldsymbol{\alpha}}\vb{x}^{\boldsymbol{\alpha}}, c_{\boldsymbol{\alpha}}\in \mathbb{K}\bigg\}
        \end{equation}
\end{definition}

\begin{definition}
The \textbf{degree of a polynomial} is the maximum of the degrees of its constituting monomials.
\end{definition}

\begin{definition}
A \textbf{term} is given by $c\vb{x}^{\boldsymbol{\alpha}}$ where $c\in\mathbb{K}$
\end{definition}

\subsection{Polynomial and Monomial Ideals}

First, what is an ideal?

\begin{definition}{(Ideal)}
 Let $\mathbb{R}$ be a ring. A subset $\mathcal{I}$ of $\mathbb{R}$ is an \textbf{ideal} if
    \begin{enumerate}[label=(\roman*)]
        \item $0_{\mathbb{R}}\in \mathcal{I}$,
        \item if $a,b\in \mathcal{I}$ then $a+b\in \mathcal{I}$, and
        \item if $a\in \mathcal{I}$ and $k\in\mathbb{R}$ then $ak = ka \in \mathcal{I}$
    \end{enumerate}
\end{definition}

\begin{definition}{(Radical of an ideal)}
The \textbf{radical of an ideal} of a ring $\mathbb{R}$, denoted by $\text{rad}(\mathcal{I})$ or $\sqrt{\mathcal{I}}$, is defined as
    \begin{equation}
        \sqrt{\mathcal{I}} \coloneqq  \Big\{a\in\mathbb{R}\text{ }\big|\text{ }a^n\in \mathcal{I} \text{ for some } n\in\mathbb{Z}^+\Big\}
    \end{equation}
\end{definition}

\begin{note}
  $\mathcal{I}\subseteq\sqrt{\mathcal{I}}$
\end{note}

Let $S$ be a set of polynomials $f$ such that $S\subseteq \mathbb{R}[\vb{x}]$, where $\mathbb{R}[\vb{x}]$ is a polynomial ring. The set $S$ corresponds to the set of equality constraints.

The system $S$ generates a \textbf{polynomial ideal} $\mathcal{I}$ of the polynomial ring $\mathbb{R}[\vb{x}]$, such that $S\subseteq \mathcal{I}$, and, for $f_1,f_2\in \mathcal{I}$, $f_1+f_2\in \mathcal{I}$; for $f\in \mathcal{I}$ and $r\in\mathbb{R}$, $rf\in \mathcal{I}$.

\textbf{Ideal generated by a set of functions.} For the polynomials $f_1,\dots,f_s\in S$, define $\expval{f_1,\dots,f_s}$ as the set that consists of all polynomials that are obtained by $\sum_{i=1}^sh_if_i$, with $h_i\in \mathbb{R}[\vb{x}]$.
\begin{equation}
    \expval{f_1,\dots,f_s} \coloneqq \bigg\{f\text{ }\Big|\text{ }f = \sum_{i=1}^sh_if_i, h_i\in\mathbb{R}[\vb{x}]\bigg\}
\end{equation}Note that $\expval{f_1,\dots,f_s}$ is an ideal, called the `ideal generated by $S$'.

\begin{theorem}{(Hilbert Basis Theorem)}
 Every polynomial ideal in $\mathbb{R}[\vb{x}]$ is finitely generated.
\end{theorem}

\begin{definition}{(Monomial ideal)}
A \textbf{monomial ideal} is a polynomial ideal that can be generated by monomials.
\end{definition}
A polynomial belongs to a monomial ideal $\mathcal{I}$ if and only if all of its terms $\in \mathcal{I}$.

\section{Duality between affine variety and coordinate ring}

First the definitions:

\begin{definition}{(Affine variety)}
The \textbf{affine variety} $\mathcal{V}(S)$ is the set of common zeros $\vb{z} = (z_1,\dots,z_n)$ of the polynomials in $S$.
    \begin{equation}
        \mathcal{V}(S) \coloneqq  \big\{\vb{z}\text{ }\big|\text{ }f(\vb{z}) = 0, \forall f\in S \big\}
    \end{equation}
\end{definition}

It’s not hard to see why we would be interested in the affine variety $\mathcal{V}(S)$. This is the set of all feasible solutions to the set of constraints $S$.

\begin{definition}{(Coordinate ring)}
Given an affine variety $\mathcal{V}$ in an affine-n space over the field $\mathbb{K}$, the \textbf{coordinate ring} of $\mathcal{V}$ is the quotient ring
\begin{equation}
    \mathbb{K}[\mathcal{V}] \coloneqq  \mathbb{K}[\vb{x}]\big/\mathcal{I}(\mathcal{V})
\end{equation}
(basically the set $\mathbb{K}[\vb{x}]\mod \mathcal{I}(\mathcal{V})$) where $\mathcal{I}(\mathcal{V})$ is the ideal formed by all polynomials $f(\vb{x})$ with coefficients in $\mathbb{K}$ which are zero at all points of $\mathcal{V}$
\end{definition}

There is a \textbf{duality} between affine algebraic varieties (i.e., the loci of common zeros of system of polynomial equations) and coordinate rings.
\begin{equation}
    {\text{affine algebraic varieties}} \sim {\text{coordinate rings}}
\end{equation}

\begin{example}
Let's look at the following example of an affine algebraic variety
\begin{equation}
    \mathcal{V} = \text{the unit circle in }\mathbb{R}^2
\end{equation}
The very same data (the set of points lying on the unit circle) is captured algebraically with the coordinate ring
\begin{align}
    &\mathbb{R}\big/\expval{x^2+y^2-1}\\
    &= \text{(polynomials in x and y) mod }(x^2+y^2-1)
\end{align}
which contains polynomials such as $(x-\cos\theta)(y-\sin\theta), \theta \in [0,2\pi)$.
\end{example}

\begin{theorem}
 $\mathcal{V}(S) = \mathcal{V}(\mathcal{I}) = \mathcal{V}(\sqrt{\mathcal{I}})$
\end{theorem}
\begin{proof}
 Let $\vb{z}_S$ be an element of $\mathcal{V}(S)$, i.e., $\forall f\in S$, $f(\vb{z}_S)=0$. Thus, for $r\in\mathbb{R}$, $rf(\vb{z}_S) = r\times0 = 0$, and for $f_1, f_2 \in S$ (hence $\in \mathcal{I}$), $f_1(\vb{z}_S)+f_2(\vb{z}_S) = 0$. Thus, each element of $\mathcal{V}(S)$ is an element of $\mathcal{V}(\mathcal{I})$. Also, since $\forall f\in S$ also $\in \mathcal{I}$, and for each element $\vb{z}_{\mathcal{I}}$ of $\mathcal{V}(\mathcal{I})$, $f(\vb{z}_{\mathcal{I}})$ needs to be $0$, which makes $\vb{z}_{\mathcal{I}}$ an element of $S$. Hence, $\mathcal{V}(S) = \mathcal{V}(\mathcal{I})$. The statement $\mathcal{V}(\mathcal{I}) = \mathcal{V}(\sqrt{\mathcal{I}})$ can be easily seen to be true.
\end{proof}
 
 The ideal $\mathcal{I}$ generated by $S$ reveals hidden polynomials. For
instance, if one of the hidden polynomials is the constant polynomial $1$
(i.e., $1\in \mathcal{I}$), then the system $S$ is inconsistent (because
$1\neq0$). To be precise, the set of all hidden polynomials is given by the
radical ideal $\sqrt{\mathcal{I}}$.  

\begin{theorem}{(Nullstellensatz)}
 $\mathcal{I}(\mathcal{V}(\mathcal{I})) = \sqrt{\mathcal{I}}$, where $\mathcal{I}(\mathcal{V}(\mathcal{I}))$ is the ideal that contains all polynomials that vanish for the variety $\mathcal{V}(\mathcal{I})$.
\end{theorem}

This duality, and the reformulation of the set $S$ so as to easily find it's common roots $\vb{z}_S$ are what we are primarily interested in. It's true that $\mathcal{V}(S) = \mathcal{V}(\mathcal{I}) = \mathcal{V}(\sqrt{\mathcal{I}})$, but $S\subseteq \mathcal{I}\subseteq\sqrt{\mathcal{I}}$ and $\sqrt{\mathcal{I}}$ is infinite. We seem to have made our problem worse. Be patient, in Sec 2.4 we shall see a finite set of functions (spoiler alert: the Gr{\"o}bner basis) which are the generator for $\sqrt{\mathcal{I}}$, and have very nice properties.

\section{Monomial Ordering}
\begin{definition}{(Monomial ordering)}
$A$ \textbf{monomial ordering} on $\mathrm{C}[\mathrm{x}]$ is a relation $\succ$ on $\mathbb{Z}_{+}^{n}$ (i.e., the monomial exponents), such that:
\begin{enumerate}[label=(\roman*)]
    \item The relation $\succ$ is a total ordering.
    \item If $\alpha \succ \beta,$ and $\gamma \in \mathbb{Z}_{+}^{n},$ then $\alpha+\gamma \succ \beta+\gamma$
    \item The relation $\succ$ is a well-ordering (every nonempty subset has a smallest element).
\end{enumerate}
\end{definition}

There are several term orderings which are of interest to us, for instance:
\begin{itemize}
    \item Lexicographic ("dictionary"). $\alpha \succ_{\text {lex }} \beta$ if the left-most nonzero entry of $\alpha-\beta$ is positive. Notice that a particular order of the variables is assumed, and by changing this, we obtain $n !$ nonequivalent lexicographic orderings.
    
    \item Graded lexicographic. Sort first by total degree, then lexicographic, i.e., $\alpha \succ_{\text {grlex }} \beta$ if $|\alpha|>|\beta|,$ or if $|\alpha|=|\beta|$ and $\alpha \succ_{\operatorname{lex}} \beta$
    
    \item Graded reverse lexicographic. $\alpha \succ_{\text {grevlex }} \beta$ if $|\alpha|>|\beta|,$ or if $|\alpha|=|\beta|$ and the right-most nonzero entry of $\alpha-\beta$ is negative. This ordering, although somewhat nonintuitive, has some desirable computational properties.
    
\end{itemize}

\begin{example}
Consider the polynomial ring $\mathbb{R}[x, y]$. In the lexicographic ordering
    $\left(\prec_{\text {lex}}\right)$ discussed above, we have:
$$
1 \prec y \prec y^{2} \prec \cdots \prec x \prec x y \prec x y^{2} \prec \cdots \prec x^{2} \prec x^{2} y \prec x^{2} y^{2} \prec \cdots.
$$

The graded lexicographic order gives $\left(\prec_{\text {grlex}}\right.$, and
the graded reverse lexicographic  order givesn $\left.\prec_{\text {grevlex}}\right)$.
In a special case consisting of two variables, the ordering coincide:
$$
1 \prec y \prec x \prec y^{2} \prec x y \prec x^{2} \prec y^{3} \prec x y^{2} \prec x^{2} y \prec x^{3} \prec \cdots.
$$
\end{example}

\begin{definition}{(Leading term)}
 Given a term order on $\mathbb{R}[\vb{x}]$ and a polynomial $f$ in this ring, say  the  highest  monomial  in $f$ with respect to the term order is $c_{\boldsymbol{\alpha}}\vb{x}^{\boldsymbol{\alpha}}$, then
 \begin{itemize}
     \item \textbf{Leading term}, $LT(f) = c_{\boldsymbol{\alpha}}\vb{x}^{\boldsymbol{\alpha}}$
     \item \textbf{Leading monomial}, $LM(f) = \vb{x}^{\boldsymbol{\alpha}}$
     \item \textbf{Leading coefficient}, $LC(f) = c_{\boldsymbol{\alpha}}$
 \end{itemize}
\end{definition}

\section{Gr{\"o}bner Basis}
Just a little more patience, we are getting to it.

\subsection{Initial Ideal}
\begin{definition}{(initial ideal)}
Consider an ideal $\mathcal{I}\subset\mathbb{R}[\vb{x}]$, and a fixed monomial ordering. The \textbf{initial ideal} of $\mathcal{I}$,
denoted $in(\mathcal{I})$, is the monomial ideal generated by the leading monomials of all the elements in $\mathcal{I}$, i.e.,
\begin{equation}
    in(\mathcal{I}) \coloneqq  \expval{LM(f)\text{ }|\text{ }f\in \mathcal{I}\backslash\{0\}}
\end{equation}
\end{definition}

Given an ideal $\mathcal{I} = \expval{f_1,\dots, f_s}$, we can construct two monomial ideals associated with it. We have the initial ideal $in(\mathcal{I})$, previously defined. We can also consider the monomial
ideal generated by the initial monomials of the generators, i.e., $\expval{LM(f_1),\dots, LM(f_s)}$. Although we always
have $\expval{LM(f_1),\dots, LM(f_s)} \subseteq in(\mathcal{I})$ and \textbf{in general these two monomial ideals are not equal}.

\subsection{Gr{\"o}bner Basis}
Finally here. As we saw in the previous section, $\expval{LM(f_1),\dots, LM(f_s)} \subseteq in(\mathcal{I})$. It is possible to produce a set of generators polynomials for which these two ideals are same. \textbf{This is the Gr{\"o}bner basis}.

\begin{definition}{(Gr{\"o}bner Basis)}
Consider the polynomial ring $\mathbb{R}[\vb{x}]$, with a fixed monomial ordering, and an ideal $\mathcal{I}$. A finite set of polynomials $\{g_1,\dots, g_s\} \subset \mathcal{I}$ is a Gr{\"o}bner basis of $\mathcal{I}$ if the $in(\mathcal{I})$ is generated by
the leading terms of the $g_i$, i.e.,
\begin{equation}
    in(\mathcal{I}) = \expval{LM(g_1),\dots,LM(g_s)}
\end{equation}
\end{definition}

\begin{theorem}
 Every ideal $\mathcal{I}$ has a Gr{\"o}bner basis $\mathcal{G}$. Furthermore, $\mathcal{I} = \expval{g_1,\dots, g_s}$.
\end{theorem}

Gr{\"o}bner bases as defined are not unique. This can be easily fixed by refining the concept to the so-called reduced Gr{\"o}bner bases, which are uniquely defined. For simplicity, we use the term `Gr{\"o}bner basis’ to refer to the reduced Gr{\"o}bner basis.

Continuing from where we left off in section 2.2, the radical ideal $\sqrt{\mathcal{I}}$ has a finite generating set, the Gr{\"o}bner basis $\mathcal{B}$, which one might take to be a triangularization of the ideal $\sqrt{\mathcal{I}}$. It is, in fact, the generalization of Gaussian elimination in linear systems.

\subsubsection*{Properties}
\begin{itemize}
    \item $\mathcal{V}(S) = \mathcal{V}(\mathcal{I}) = \mathcal{V}(\sqrt{\mathcal{I}}) = \boxed{\mathcal{V}(\mathcal{B})}$
    
    Hence, the set of all feasible solutions of $S$ is the same as the set of solutions for $\mathcal{B}$.

    \item If $1\in\mathcal{B}$ then $\mathcal{V}(S) = \phi$
    
    Hence, it can be quickly figured out whether there exists a feasible solution to the constraints defining $S$. Other algorithms to solve IP problems take very long if there is no feasible solution. Gr{\"o}bner basis computation takes a long time when there are a lot of feasible solutions, in which case the other algorithms are faster. So if we run Gr{\"o}bner basis computation in parallel to some other algorithm, we reach an answer much quickly one way or the other.

    \item The size of $\mathcal{V}(S) = \mathcal{V}(\mathcal{B})$ can be obtained using staircase diagram (without solving equations).
\end{itemize}

\begin{figure}[H]
    \centering
    \includegraphics[width = 5.5in]{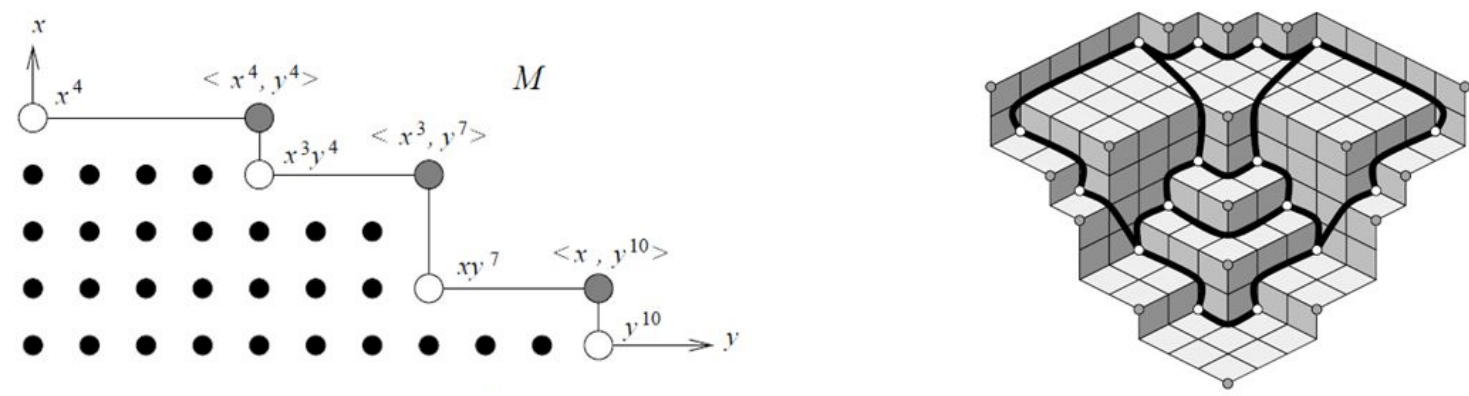}
    \caption{The monomial ideal $M = \expval{x^4, x^3y^4, xy^7, y^{10}}$, with its generators (white circles), standard monomials (black dots), and irreducible components (shaded circles)~\citep{dridi2018novel}}
    \label{fig:my_labetsl}
\end{figure}

\begin{example}
Consider the system given by,
\begin{equation}
    \mathcal{S} = \{x^2 + y^2 + z^2 - 4 = 0, x^2 + 2y^2 - 5, xz - 1 = 0\}
\end{equation}

We want to solve $\mathcal{S}$. One way to do so is to compute a Gr{\"o}bner basis for $\mathcal{S}$. We can do this directly by using the Sympy\footnote{www.sympy.org/} package in Python, which uses the Buchberger’s algorithm. It was specified to use the lexicographic ordering such that $x > y > z$. The computed Gr{\"o}bner basis is,
\begin{equation}
    \mathcal{B} = \{x+2z^3-3z, y^2-z^2-1, 2z^4-3z^2+1\}
\end{equation}
We can see that the initial system has been triangulized: The last equation contains only the variable $z$, while the second has an additional variable, and so on. The variable z is said to be eliminated with respect to the rest of the variables. For the defined lexicographic ordering, the program tries to isolate $z$ first, then $z$ and $y$, and finally $x$, $y$, and $z$ (all variables). It is clear that different orderings yield different Gr{\"o}bner bases.
\end{example}

\subsection{How to compute the Gr{\"o}bner basis?}

The following three definitions are useful in defining the Buchberger algorithm, which computes the Gr{\"o}bner basis.

\begin{definition}{(S-Polynomials)}
Given two multivariate polynomials $f,g\in\mathbb{R}[\vb{x}]$, considering the least common multiple of their leading monomials with respect to an ordering $\succ$,
\begin{equation}
    L = \text{lcm}(LM(f),LM(g))
\end{equation}
we define, the \textbf{S-Polynomial} as,
\begin{equation}
    S(f,g) = \frac{L}{LT(f)}f - \frac{L}{LT(g)}g
\end{equation}
\end{definition}

\begin{note}
  $S(f,g)$ is defined this way so that the leading terms of $f,g$ cancel out.
\end{note}

\begin{definition}{(Normal or reduced form)}
A polynomial $f\in\mathbb{R}[\vb{x}]$ is \textbf{reduced} or \textbf{normal} with respect to $G = \{g_1,\dots,g_s\}\subset\mathbb{R}[\vb{x}]$ if no monomial of $f$ is contained in the ideal $\expval{LM(g_1),\dots,LM(g_s)}$, i.e., no term in $f$ is a multiple of any $LM(g_i)$.
\end{definition}

Another formal definition of Gr{\"o}bner Basis:
\begin{definition}
For all the pairwise elements of the Gr{\"o}bner basis $\mathcal{G}$,
\begin{equation}
    \forall g_i, g_j\in\mathcal{G}, \mod(S(g_i,g_j),\mathcal{G}) = 0
\end{equation}
\end{definition}

\subsection*{Buchberger algorithm}
\textbf{Input:} A set of polynomials $F$ that generates an ideal $\mathcal{I}$\\\\
\textbf{Output:} A Gr{\"o}bner basis $\mathcal{G}$ for $\mathcal{I}$\\\\
\textbf{Algorithm:}

\begin{adjustwidth}{0.3in}{}
Initialize $\mathcal{G} \coloneqq F$\\
And a set, $M \coloneqq \{\{f_i,f_j\}\text{ }|\text{ }f_i,f_j\in\mathcal{G} \And f_i\neq f_j\}$\\\\
while $M\neq\phi$
    \begin{adjustwidth}{0.3in}{}
    $\{p,q\} \coloneqq \text{ an element of }M$\\
    Update $M$, $M \coloneqq M\backslash\{p,q\}$\\\\
    Define the S-polynomial, $S = S(p,q)$\\
    Reduce $S$ with respect to $\mathcal{G}$, $h \coloneqq \text{reduced}(S,\mathcal{G})$\\\\
    if $h\neq0$:
        \begin{adjustwidth}{0.3in}{}
        Add the new pairs $\{g,h\}$ to $M$, $M \coloneqq M\cup\{g,h\}\forall g\in\mathcal{G}$\\
        Add $h$ to the set $\mathcal{G}$, $\mathcal{G} = \mathcal{G}\cup h$ (corresponding to which we have already updated $M$)
        \end{adjustwidth}
    \end{adjustwidth}
\end{adjustwidth}

The output of the Buchberger algorithm may not be the reduced Gr{\"o}bner basis.
\begin{definition}{(Reduced Gr{\"o}bner basis)}
A Gr{\"o}bner basis $\mathcal{G}$ is reduced if $\forall g\in\mathcal{G}$
\begin{enumerate}[label=(\roman*)]
    \item $LT(g)$ is not a factor of any element of $\mathcal{G}\backslash\{g\}$
    \item $LC(g) = 1$
\end{enumerate}
\end{definition}

\subsubsection*{Special cases of Buchberger algorithm:}
\begin{itemize}
    \item If the system of equations is univariate, it reduces to Euclid’s division algorithm
    \item If the system of equations is linear it reduces to Gauss’ algorithm
\end{itemize}

\subsection{Applications of Gr{\"o}bner basis}
\begin{itemize}
    \item Algebraic Geometry
    \item Coding Theory
    \item Cryptography
    \item Invariant Theory
    \item \textbf{Integer Programming}
    \item \textbf{Graph Theory}
    \item Statistics
    \item Symbolic Integration
    \item Symbolic Summation
    \item Differential Equations
    \item Systems Theory
\end{itemize}

\section{Example: Graph coloring using Gr{\"o}bner basis}
Given a graph $\mathcal{G}(V, E)$, where $V$ is the set of vertices and $E$ is the set of edges, given $k$ colors, $K = \{1,\dots,k\}$,
\begin{itemize}
    \item Can the graph be colored s.t. no two vertices connected by an edge have the same color?
    \item If yes, what are the possible colorings?
\end{itemize}
This can be posed as an IP (actually as a SAT since the objective is irrelevant):
\begin{align}
    &\hspace{0.2in}\min_{\vb{x}} 1\hspace{0.1in}&&\\
    &s.t.&&\notag\\
    &\hspace{0.2in}x_{ij}\in\{0,1\},\forall j\in K,\forall i\in V && (\text{If V is colored j, $x_{ij} = 1$, else $0$})\\
    &\hspace{0.2in}\sum_{j\in K}x_{ij} = 1, \forall i\in V && (\text{each vertex must have one (and only one) color})\\
    &\hspace{0.2in}x_{uj}+x_{vj}\leq1,\forall j\in K, \forall (u,v) \in E && (\text{Either none or only 1 of the 2 $V$'s of an $E$ have a color $j$})
\end{align}
The constraints can be remodelled in a way which is more suitable for the Gr{\"o}bner basis computation as follows. We map the $k$ colors to the $k$-th roots of unity, i.e., for $\zeta = \exp(\iota\frac{2\pi}{k})$, we map the colors $\{1,\dots,k\}$ to $\{1,\zeta,\dots,\zeta^{k-1}\}$. Thus, the statement that every vertex must have one of the colors can be written as the set of polynomial equations:
\begin{equation}
    \mathcal{S}_k = \{ x_i^k - 1 = 0 : i = 1, 2, \ldots, n \}
\end{equation}
We also require that two adjacent vertices $x_i$ and $x_j$ are assigned different colors. From the previous discussion we know that $x_i^k = 1$ and $x_j^k = 1$, so $x_i^k = x_j^k$ or, equivalently, $x_i^k - x_j^k = 0$. By factorization we obtain that:
$$x_i^k - x_j^k = (x_i - x_j) \cdot f(x_i, x_j) = 0$$
where $f(x_i, x_j)$ is a bivariate polynomial of degree $k-1$ in both variables. Since we require that $x_i \not= x_j$ then $x_i^k - x_j^k$ can vanish only when $f(x_i, x_j) = 0$. This allows us to write another set of polynomial equations:
$$\mathcal{S}_{\mathcal{G}} = \{ f(x_i, x_j) = 0 : (i, j) \in E \}$$
Next we combine $\mathcal{S}_k$ and $\mathcal{S}_{\mathcal{G}}$ into one system of equations $\mathcal{S}$. The graph $\mathcal{G}(V, E)$ is $k$-colorable if the Gröbner basis $\mathcal{B}$ of $\mathcal{S}$ is non-trivial, i.e., $\mathcal{B} \not= \{1\}$. If this is not the case, then the graph isn’t $k$–colorable. Otherwise the Gröbner basis gives us information about all possible $k$–colorings of $\mathcal{G}$.

\section{Algebraic Geometry in Combinatorial Optimization}

\subsection*{The BPT method}
\vspace{-0.2in}
Proposed by~\citet{bertsimas2000new}.

Given an optimization problem,
\begin{align}
    &\text{argmin}_{\mathbf{x}} f(\mathbf{x})\\
    &\text{s.t. }g(\mathbf{x})=0 \\
    &\mathbf{x} \in\{0,1\}^{n}
\end{align}
where $f(\mathbf{x}), g(\mathbf{x}) \in \mathbb{R}[\mathbf{x}]$, Algebraic Geometry appears naturally.

\begin{itemize}
\item The feasible solutions are the variety of the ideal generated by the constraints, additional polynomials whose roots are $\{0,1\}$, enforcing the variables to be binary: $x_i(x_1-1) = x_{i}^{2}-x_{i}$ and a polynomial $z-f(\vb{x})$ with a new variable $z$. The ideal is given by,
$$
\mathcal{I}=\left\langle z-f(\mathbf{x}), g(\mathbf{x}), x_{i}^{2}-x_{i}\right\rangle \subset \mathbb{R}\left[z, x_{0}, \cdots, x_{n-1}\right]
$$
\item We compute it's Gr{\"o}bner basis $\mathcal{B}$ with the lexicographic ordering $x_i > z$, which results in $\mathcal{B}$ containing a polynomial $g(z)$ which is just a function of $z$.

\item Solving $g(z) = 0$ gives the set of all feasible values $z_{sol} = \{z_1,\dots,z_m\}$ of the objective function $f(\vb{x})$.
\item Choose the optimal value of $z_{sol}$, $z_{opt} = \min(z_{sol})$
\item Solve rest of the polynomials in $\mathcal{B}$ by substituting $z = z_{opt}$, which gives the solution $\vb{x}_{sol} = \vb{x}_{opt}$
\end{itemize}

Hence, we have obtained the solution $\vb{x}_{opt}$ to the optimization problem.

This ends Part 1 of Lecture 2.

\lecture{3}
{Test-set methods - Gr{\"o}bner Basis (Part 2)}
{Prof. Sridhar Tayur, David E. Bernal}
{Srijan Gupta, Vighnesh Natarajan}
{https://bernalde.github.io/QuIP/slides/47-779\%20Lecture\%202\%20-\%20Test-set\%20methods.pdf}
{https://colab.research.google.com/github/bernalde/QuIP/blob/master/notebooks/Notebook\%202\%20-\%20Groebner\%20basis.ipynb}
{https://cmu.zoom.us/rec/play/OHWbc7V9tJKIeKrzmGE-Ms0WQQlXAQEp3SLKBnlFP5nJEGfKFJPE4vxU13M-PWZYS4pCGbh9_AJ-RNw.pOCUff3tcAcKq-_D}
{RV=9U^s3}

\section{Conti and Traverso method}
The Conti and Traverso method~\citep{conti1991buchberger} or the CT method is a purely algebraic method of solving integer programs. This method is an application of the Buchberger algorithm to integer programs. We want to solve an IP problem with the following constraints.
\begin{align*}
    \underset{x}{\text{minimize}}\; & c^T x\\
    \text{subject to:} \qquad & Ax = b, x \in \mathbb{Z}^n\\
\end{align*}
$A$ is the coefficient matrix, $A \in \mathbb{Z}^{m \times n}_+$, b is the right hand side vector, $b \in \mathbb{Z}^{m}_+$ and $c$ is the cost vector with $c \in \mathbb{Z}^{n}_+$. Here, we must note that (1) b must be in the column space of A for a feasible solution; (2) the constraint Ax = b gives us a polytope. We create a term ordering with respect to an inner product with respect to the cost vector $c$.
\paragraph{Toric Ideal} A Toric ideal is a mapping of fields generated by the difference of monomials. For a mapping  $\mathbb{K}^2 \; \{x, y\} \longrightarrow \mathbb{K}^3 \; \{x^2, xy, y^2\}$, the Toric ideal is $<z_1z_3 - z_2^2>$. There is a mapping to a variable z, to which we shall also map the integer problem constraints.\\\\
The first step is to create the Toric ideal of the constraints. The constraints of Ax = b are mapped to a similar variable z, of a Toric ideal, which gives us, for each constraint
\begin{align*}
    z_i^{a_{i1}x_1+a_{i2}x_2....a_{in}x_n} = z_i^{b_i}
\end{align*}
Thus, for all constraints, we get,
\begin{align*}
    \prod_{i=1}^m \prod_{j=1}^n (z_i^{a_{ij}x_j}) = \prod_{i=1}^n z^{b_i}
\end{align*}
With this, we define a mapping $\phi: \mathbb{Q}[w_1,...w_n] \longrightarrow \mathbb{Q}[z_1,...z_m, z_1^{-1}...z_m^{-1}]$ such that $\phi(w_j) = \prod_{j=1}^n (z_i^{a_{ij}x_j})$. We can write the columns of the matrix A as $a_j = a_j^+ - a_j^-$. This lets us define our ideal, 
\begin{align*}
    \mathcal{I} =\; <z^{a_j^-}w_j - z^{a_j^+}, 1 - z_1 - ... -z_m>
\end{align*}
Before proceeding further, let us look at a couple of examples for computing the ideal.
\paragraph{Example 1}
\begin{align*}
    4x_1 + 5x_2 + x_3 & = 37\\
    2x_1 + 3x_2 + x_4 & = 20
\end{align*}
It is clear that $\phi(w_1) = z_1^4z_2^2$, $\phi(w_2) = z_1^5z_2^3$, $\phi(w_3) = z_1$, $\phi(w_4) = z_2$. A set of feasible solutions satisfy $\phi(w_1, w_2, w_3, w_4) = z_1^{37}z_2^{20}$. Let $f_i = \phi(w_i)$. Our ideal of consideration becomes
\begin{align*}
    \mathcal{I} =\; <f_1-w_1,f_2-w_2,f_3-w_3,f_4-w_4>
\end{align*}
\paragraph{Example 2} We can also consider negative coefficients in the constraints.
\begin{align*}
    2x_1 - 1x_2 + x_3 & = 4\\
    -1x_1 + 2x_2 & = 5
\end{align*}
It is clear that $\phi(w_1) = z_1^2/z_2$, $\phi(w_2) = z_2^2/z_1$, $\phi(w_3) = z_1$. A set of feasible solutions satisfy $\phi(w_1, w_2, w_3) = z_1^{4}z_2^{5}$. Our ideal in this case becomes
\begin{align*}
    \mathcal{I} =\; <w_1z_2 - z_1^2,w_2z_1 - z_2^2,w_3-z_1>
\end{align*}

\noindent Considering $\mathcal{I} = \langle z^{a_j^-}w_j - z^{a_j^+}, 1 - z_1 - ... -z_m \rangle = \langle x^u - x^v: Au = Av, u, v \in \mathbb{Z}^n_+ \rangle$, we define a monomial relative ordering with respect to the cost c, $>_c$, defined as
\begin{align*}
    \alpha >_c \beta \longrightarrow \;\;& c^T\alpha > c^T\beta \;\;or\\
    & c^T\alpha = c^T\beta\;\; and\;\; \alpha > \beta
\end{align*}
With a Gr{\"o}bner basis with respect to monomial ordering, we get $\mathcal{B}\cap\mathbb{Q}[w]$ is a Gr{\"o}bner basis of the ideal $\mathcal{I}\cap\mathbb{Q}[w] = \mathcal{I}_A$. $\mathcal{I}_A$ is called the Toric ideal of A. We can observe that this is independent of the right hand side of the constraints. Thus, for a problem, if we can find the Gr{\"o}bner basis of the constraint matrix $A$, then no matter what the right hand side is, we can quickly solve the problem. 
\paragraph{Theorem} Let $\mathcal{B}_{>_c}$ be the reduced Gr{\"o}bner basis with respect to $>_c$ of the Toric ideal $\mathcal{I}_A$. Then for any right-hand side and integer constraint matrix, term order, and non-optimal feasible solution $z_0$, there is some pair $u,v$ such that $z_1 = z_0 - u + v$ is a better feasible solution. \\\\
We can use this to generate the reduced Gr{\"o}bner basis either by serial reduction or by a direct reduction using python. Using the Gr{\"o}bner basis, we get our solution of the Integer program. A link to the examples on python has been added in the references.

\section{Improvements and Benchmarks in computing Gr{\"o}bner basis}
The main step in these algorithms' performance is the computation of the Gr{\"o}bner basis. Sympy takes around 15 seconds for a 12 vertex graph coloring problem. Mathematica is more optimized, giving the answer within a second. As the size of the problem grows, we can see that we will get stuck. 

There have been various improvements optimizing the Buchberger algorithm. It has a doubly exponential time complexity and is a crucial step in optimization. Besides, there are optimizations called project and lift, which optimize the basis vectors' serial reduction. Optimizations by Charles Faugere are based on a transformation into linear algebraic problems, where more efficient algorithms can be applied. For the efficient computation of the Gr{\"o}bner basis of the Toric ideals, efficient algorithms try to avoid the zero reduction of the polynomial during back substitution. 4ti2 is the most efficient algorithm to compute this.

\noindent As constraints get complicated, we can't simply apply the BPT and the CT(Conti Traverso) methods on the entire formulation. Techniques used for complicated constraints are first to relax the problem to a linear problem and perform a \textbf{walk back} from the solution. This applies to all branch and bound solvers. A key critical factor is that we always need to stay feasible to find the solution. Finding the Gr{\"o}bner basis is the best way to guarantee feasibility. This way lets you iterate through a lattice of feasible points. This leads us to the concepts of test sets. 

\section{Test-set methods}
Given an integer linear program, there exists a finite set called the test set denoted by $\mathcal{T} = \{t_1, t_2....t_n\}$ that only depends on A and c, the constraint matrix and the cost vector, that assures a feasible solution $x^*$ is optimal if and only if $c^T(x^*+t_i) \geq c^Tx^*$ whenever $x^*+t_i$ is feasible. The essential idea is to create a separation of constraints, calculate the Gr{\"o}bner basis, and do a walk back with the complex constraints acting as a complicated oracle.\\

\noindent A clever observation is that the Gr{\"o}bner basis of a Toric ideal is a test set. Hence we can see that the CT method is a test set method. However, things can get more complicated for general constraints, and we will need to perform walk backs for better optimization.

The following is a comparison between the classic divide and conquer methods and test set-based methods.
\begin{multicols}{2}
{\bf Methods based on divide-and-conquer}
\begin{itemize}
    \item Branch-and-Bound algorithms
    \item Harness advances in polyhedral theory
    \item Global optimality guarantees
    \item Efficient codes and exponential complexity
\end{itemize}

\columnbreak
{\bf Methods based on test-sets}
\begin{itemize}
    \item Algorithms based on “augmentation”
    \item Harness tools from algebraic geometry
    \item Global convergence guarantees
    \item Few implementations available, Polynomial oracle complexity once we have test-set
\end{itemize}
\end{multicols}

The requirements to work with test sets on an Integer program are 
\begin{itemize}
    \item An initial feasible solution
    \item An oracle to compare the objective function
    \item A test set (set of directions)
\end{itemize}
Given an objective, the test set will give us directions to improve the objective. If we cannot improve the objective, we have reached optimality. The Gr{\"o}bner basis test-set only depends on the constraints and objective and can be computed for equality constraints with integer variables.

\paragraph{Example}
\begin{align*}
    Ax = \begin{bmatrix}
1 & 1 & 1 & 1\\
0 & 1 & 2 & 3
\end{bmatrix}x = \begin{bmatrix}
10\\15
\end{bmatrix}
\end{align*}

\begin{figure}[htb]
\centering
\includegraphics[scale=0.4]{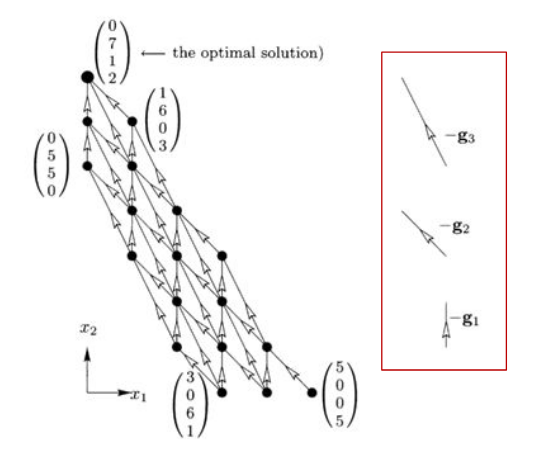}   
\caption{2D slice of a 4D lattice. Improvement by following directions of the test set to optimality}
\label{fig:test}
\end{figure}
\noindent Figure~\ref{fig:test} is a 2D section of a 4D lattice. From a feasible point, we are using the test set to iterate to optimality. The test set direction vectors g1, g2 and g3 have been obtained by finding the Gr{\"o}bner basis of the allied problem's objective and constraint matrix. The computation of this test set is the NP-hard part. The rest of the problem is polynomial, as discussed above.\\

\noindent Chance constraints are a particularly difficult constraint to deal with. This is tackled by relaxing these constraints to find a gr{\"o}bner basis and using the chance constraints as a complex oracle during the walk back. Regular MINLP solvers fail on chance constraints. However, chance constraints can be reformulated and tackled using ILP solvers. However, these ILP solvers struggle and are only competitive in verifying optimality. 

\begin{figure}[htb]
\centering
\includegraphics[scale=0.4]{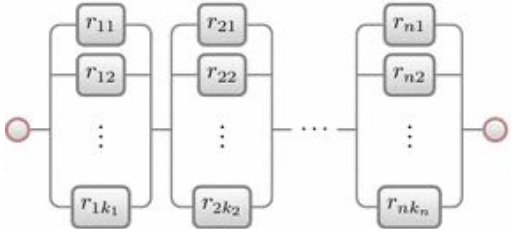}   
\caption{Series parallel process~\citep{gago2015improved}}
\label{fig:process}
\end{figure}

\paragraph{Example} Process reliability\\
We have a series-parallel process, shown in Fig.~\ref{fig:process}, and we want to minimize cost, while satisfying a reliability threshold.
\noindent With any unit failing, we want to ensure the entire system does not fail with a certain reliability. Hence, back-up units are added, which take up the cost, but also improve reliability. The formulation of this problem is
\begin{align*}
    \underset{x}{\text{minimize}} \qquad & \Sigma_{i,j} c_{ij}x_{ij}\\
    \text{subject to:} \qquad & 0 \leq x_{ij} \leq u_{ij}\\
    & \Sigma_j x_{ij} \geq 1\\
    & x_{ij} \in \mathbb{Z}\\
     \prod_{i=1}^n (1-\prod_{j=1}^{k_j}(1-r_{ij})^x_{ij}) & = R(x) \geq R_o
\end{align*}
Here, $c_{ij}$ is the cost of unit $i,j$, $x_{ij}$ is the number of backups of a particular unit, $u_{ij}$ is the maximum number of each unit, $r_ij$ is the reliability of a unit, and $R_o$ is the reliability threshold. This problem is relaxed to an LP 
\begin{align*}
    \underset{x}{\text{minimize}} \qquad & \Sigma_{i,j} c_{ij}x_{ij}\\
    \text{subject to:} \qquad &  x_{ij} + t_{ij} = u_{ij}\\
    & \Sigma_j x_{ij} - d_{ij} = 1\\
    & x_{ij} \in \mathbb{Z}, \qquad i \in I, j \in J\\
    & \Sigma_{i}^n \; \Sigma_j^{k_j} \; c_{ij} x_{ij} = b
\end{align*}

\noindent An interesting part about this problem is that the Gr{\"o}bner basis can be calculated analytically.
\begin{align*}
    \mathcal{B}_{<_c} = \{x_{ik}d_i - t_{ik}b^{c_{ik}}, x_{iq}t_{ip} - x_{ip}t_{iq}b^{c_{iq}-c_{ip}} \}
\end{align*}

\lecture{4}{Graver Basis}{Prof. Sridhar Tayur, David E. Bernal}{Parth S. Shah}
{https://bernalde.github.io/QuIP/slides/47-779\%20Lecture\%203\%20-\%20Graver\%20basis.pdf}
{https://colab.research.google.com/github/bernalde/QuIP/blob/master/notebooks/Notebook\%203\%20-\%20Graver\%20basis.ipynb}
{https://cmu.zoom.us/rec/share/4QTXDhnVQag7dN0IfrOnbH16yhr6G0_5rBV3Nj3WbO1m3t4VLwqZzutHjXKlg3Bw.x4SMdIgGN8iyt9JD}
{49\&WKKAR}

\section{Introduction}
This lecture introduces the concept of Graver basis, and its application to the field of Integer programming as a test set. Rather than a rigorous mathematical definition, we focus
on a somewhat intuitive understanding of Grave basis.

\section{History}
The original paper on the Graver basis was written in 1975 by Jack E. Graver \cite{graver1}. Later, applications of Graver basis to more general objectives in integer programming were  demonstrated by Onn in 2010 \cite{graver2}. Recently there has been increasing interest in this method of solving integer programs, which inspired~\cite{GAMA}.

\section{Hilbert Basis}
Before getting into Graver basis, it is useful to first understand the Hilbert basis. This other basis will be used later to describe the Graver Basis.\newline
There are 2 methods of understanding the Hilbert Basis:
\begin{itemize}
    \item {\bf First method}: Given a cone $\mathcal{C}, \mathcal{F}$ is defined as the points within the cone with integral co-ordinates, i.e., $\mathcal{F} = \mathcal{C}$ $\cap$ $\mathbb{Z} ^{n}$. Now, $\mathcal{H}$ is considered an integral generating set of $\mathcal{F}$ if for every \textbf{x} $\in $ $\mathcal{F}$, there exists:
    \begin{equation}\notag
        \{\mathbf{h_{1},..,h_{n}}\} \mid \textbf{x} = \sum^{n}_{i=1} \lambda_{i} \mathbf{h_{i}} \; , \lambda_{i}\in \mathcal{Z}_{+}.
    \end{equation}
    If such a set of $h_{i}$ is minimal with respect to inclusion, then this set is an integral basis. If this set is finite, then it forms a \textbf{Hilbert basis}.
    \item {\bf Second method}: Considers solving a homogeneous system of linear equations like $\mathbf{Ax=b}$. The form
    of constraint here are the same as those used in our previous lecture on integer programming. 
   We want to find the kernel of the matrix $\mathbf{A}$:
    \begin{equation}\notag
        ker(A) = \{\mathbf{x} \in \mathcal{Z}_{+}^{n} \mid \mathbf{Ax=0}\}.
    \end{equation}
    The \textbf{Hilbert basis} of the matrix $\mathbf{A}$
    is the subset of the above kernel, where the kernel
    is minimal with respect to inclusion (non-zero is implicit from the definition of the kernel).
    
\end{itemize}
From a more geometric point of view, the Hilbert basis is pretty much like a lattice basis for $\mathcal{F}$. It is the set of (integral co-ordinates) vectors in $\mathcal{F}$ whose linear sums can give any vector in the space. A good animation of the same is provided in the video lecture as well~\cite{quip}.\newline
In the past~(see for instance~\cite{integral-basis}), integral basis have been used to solve integer programs using pivoting methods. Such methods differ from branch and cut, which is used today. However, there is potential to bring integral basis methods into the present day. These integral basis
method are very efficient in verifying optimality. However, the test-set (introduced in lecture 2 and required to test optimality) may be very large and hard to calculate classically.
\section{Graver Basis}
Now we can understand the basic idea of a Graver basis. Let $\mathcal{O}_{j}$ be the $j^{th}$ orthant of a $\mathcal{R}^{n}$ vector space.  Consider a constraint matrix \textbf{A} defined in the same space. Let:
\begin{equation}\notag
    \mathcal{H}_{j}(\mathbf{A}) = \mathcal{H}(\mathbf{A}) \cap \mathcal{O}_{j}
\end{equation}
Given this, the Graver basis is defined as:
\begin{equation}\notag
    \mathcal{G}(\mathbf{A}) = \bigcup_{j}\mathcal{H}_{j}(\mathbf{A})
\end{equation}

Again, the geometric intuition here is that we consider the Hilbert basis over each of the orthants (every orthant is a cone). The union of those gives us the Graver Basis \cite{quip}.

\subsection{Some notation}
\begin{itemize}
    \item Recollect how we had an ordering for the Gr{\"o}bner basis in lecture 2. Similarly, we need to define an ordering for the Graver basis as well.
    \item $\mathbf{u,v} \in \mathbb{R}^{n}$ are conformal if both the vectors belong to the same orthant, i.e. $u_{j}v_{j}  \geq 0 \; \forall \; j = \{1,2,..,n\}$
    \item For $\mathbf{u,v} \in \mathcal{R}^{n}$, $\mathbf{u} \sqsubseteq \mathbf{v}$ if $\mathbf{u}$ and $\mathbf{v}$ are conformal and $\mathbf{v}$ is farther from the origin than $\mathbf{u}$
    \item $\mathcal{L}$($\mathbf{A}$) is basically the kernel of the matrix $\mathbf{A}$. This has the property that any element $\mathbf{z}$ belonging to it has a $\sqsubseteq$ representation in terms of the Graver basis:
    \begin{equation}\notag
        \mathbf{z} = \sum_{i} \alpha_{i}g_{i}, \;\; \alpha_{i} \in \mathbb{Z}_{+}, g_{i} \in \mathcal{G(A)}, g_{i}\sqsubseteq \mathbf{z}
    \end{equation}
\end{itemize}

\subsection{Graver and Gr{\"o}bner basis}
\begin{itemize}
    \item It turns out that the Gr{\"o}bner basis and the Graver basis can be related in some way. Given an objective $\mathbf{c} \in \mathbb{Z}^{n}$ and a toric ideal $\mathcal{I}$ for the constraint matrix $\mathbf{A}$, we denote the Gr{\"o}bner basis as $\mathcal{B}\mathbf{(A,c)}$
    \item We also define the \textbf{Universal Gr{\"o}bner basis} as the union over the set of all objective functions as:
    \begin{equation}\notag
        \mathcal{U}(\mathbf{A})=\bigcup_{c\in\mathbb{Z}^n}\mathcal{B}\mathbf{(A,c)}
    \end{equation}
    \item Now it turns out that this Universal Gr{\"o}bner basis is contained in the Graver basis upto a negating vector, i.e.,
    \begin{equation}\notag
        \mathcal{U}(\mathbf{A}) \subseteq \mathcal{G}(\mathbf{A})
    \end{equation}
    \item Sometimes these two are the same, specifically in cases where the matrix $\mathbf{A}$ is totally unimodular, or it is a Lawrence lifting matrix. A totally unimodular matrix is a square matrix with every entry in the matrix: 0, -1, or +1, and the determinant being -1 or 1. A unimodular matrix just does not have a restriction on the entries except that they are integers.
\end{itemize}

\subsection{Usage of the relation between the Graver and Gr{\"o}bner basis}
Using the relation shown in the previous subsection, there is a method to find the Graver basis for a matrix by computing its Gr{\"o}bner basis instead. What is done is that the matrix is "Lawrence lifted". Mathematically what we do is consider the matrix $\mathbf{\Lambda}$ made from the original constraint matrix $\mathbf{A}$ as:
\begin{equation}\notag
    \mathbf{\Lambda}(\mathbf{A}) = 
    \begin{bmatrix}
        \mathbf{A} & 0 \\
        \mathbf{I}_n & \mathbf{I}_n
    \end{bmatrix}
\end{equation}
where $\mathbf{I}_n$ is the Identity matrix of the appropriate order. Its toric ideal (described in the previous lecture) is given by $\mathcal{I}_{\mathbf{\Lambda(A)}} = x^{u^{+}}y^{u^{-}}-x^{u^{-}}y^{u^{+}}$ where $\mathbf{u} \in \mathcal{L}(\mathbf{A})$. The $y$ terms actually come from the identity matrices. 

Interestingly, for such a matrix we have:
\begin{equation}\notag
    \mathcal{G}(\mathbf{\Lambda(A)}) = \mathcal{U}(\mathbf{\Lambda(A)})  = \mathcal{B}(\mathbf{\Lambda(A), c})
\end{equation}
for any ordering $\mathbf{c}$ defined on the vector space spanned by $\mathbf{x,y}$. To compute the Graver basis, one simply computes the Gr{\"o}bner basis for this new matrix $\mathbf{\Lambda}$ where $y_i$ is now 1(remember this comes from the identity matrices).

Code for the above algorithm can be seen in the first part of \href{https://colab.research.google.com/github/bernalde/QuIP/blob/master/notebooks/Notebook\%203\%20-\%20Graver\%20basis.ipynb}{this notebook}. The code uses the CT method described in lecture 2 for computing the Gr{\"o}bner basis. As we saw in the previous lecture, calculating the entire Gr{\"o}bner basis for a large matrix is very hard to do efficiently on a classical machine. Hence we look at other algorithms.

\subsection{Normal form algorithm}
The normal form algorithm returns the "normal form" of a vector and a set belonging to the same Lattice. The normal form $\mathbf{r}$ of $\mathbf{s} \in \mathcal{L}$ and a set $\mathcal{G} \subset \mathcal{L}$ is related to them as:
\begin{equation}\notag
    \mathbf{s} = \sum_{i} \alpha_{i}g_{i} + \mathbf{r}, \;\; \alpha_{i} \in \mathbb{Z}_{+}, \;\;g_{i} \in \mathcal{G},\;\; \mathbf{s}\sqsubseteq \mathbf{r},\;\;  g_{i} \not\sqsubseteq \mathbf{r}\;\; \forall g_{i} \in \mathcal{L}
\end{equation}
Basically we try to to find "residual" or remainder of the vector with respect to the set, and then sift through this remainder to find elements that won't leave any. The algorithm can is described below, but note that it very heavy computationally:

\begin{addmargin}[1em]{0em}
\textbf{Input}:\textbf{s, $\mathcal{G}$}\newline
\textbf{Output}:\textbf{r}\newline
\textbf{Initialise}: $\mathbf{s} \mapsto \mathbf{r}$\newline
\textbf{Loop while} $\exists \mathbf{g} \in \mathcal{G} \mid \mathbf{g} \sqsubseteq \mathbf{r}$: \newline 
\textbf{Do}: $\mathbf{r}$-$\mathbf{g}$ $\mapsto$ $\mathbf{r}$\newline
\textbf{End of loop}\newline
\textbf{Return: r}
\end{addmargin}

 This is not an algorithm to get the Graver basis, but rather helps in the understanding and implementation of the following algorithm.\newpage
 \subsection{Pottier's Algorithm}
 What would happen if the Normal Form algorithm's input is the kernel of a matrix? We would get the Graver basis! Why? Because it returns the $\sqsubseteq$-minimal elements of the set. Hence, we arrive at the following algorithm:\newline
 
\begin{addmargin}[1em]{0em}
\textbf{Input}:Generating initial set $\mathcal{F}\subseteq\mathcal{L}$ = ker(A)\newline
\textbf{Output}:Graver basis $\mathcal{G}$ of $\mathcal{L}$\newline
\textbf{Initialise}: \newline
    $\mathcal{G}$ = $\mathcal{F}\cup\mathcal{(-F)}$\newline
    $\mathcal{S} = \cup_{f,g \in \mathcal{G}} {f+g}$\newline
\textbf{Loop while} $\mathcal{C} \neq \emptyset$: \newline 
\textbf{Do}:  $\forall s \in \mathcal{C}:\;\; r \leftarrow normalform(s, \mathcal{G}) \; and \; \mathcal{C} \leftarrow \mathcal{C}| \{s\}$\newline
\textbf{if} r $\neq$ 0 \textbf{then} \newline
$\mathcal{G} \leftarrow \mathcal{G}\cup\{r\}\; and\; \mathcal{C}\leftarrow \mathcal{C}\cup \{r+g_{i}\} \; \forall g_{i} \in \mathcal{G}$ \newline
\textbf{End of if}\newline
\textbf{End of loop}\newline
\textbf{Return:} $\mathcal{G}$
\end{addmargin}

This is a sort of completion algorithm and is pretty hard to do classically. Intuitively what we are doing is that we are taking each element in the kernel and finding the residuals again. An element is added to the Graver basis, only when
the element has no residual. The main issue here is that the kernel may contain many elements that are not $\sqsubseteq$-minimal, and hence we end up spending a lot of unnecessary time on such elements.\newline
There is an improvement in this algorithm. The computation is not done on the entire set, but only a subset that conforms to certain restrictions. We don't need to take every element's sum with every other element as shown above, but only if two vectors have the same pattern in the signs of their components do we consider their sum. This is called the project and lift algorithm.\newline
Note that the most efficient implementation of this algorithm is in a software named \href{}{4ti2}.

\subsection{Graver Basis as a Test Set}
Test sets were discussed in the previous lecture as well, although there we looked at how the Gr{\"o}bner basis may be used as a test set for integer programming. Here we look at how the Graver basis works as a test set and for what kinds of objective functions. A quick recap of test sets:
\begin{itemize}
    \item Assume we are given an integer linear program:
    \begin{equation}\notag
        \min_{x}f(x) \mid \mathbf{Ax = b, l\leq x \leq u, x \in \mathbb{Z}^{n}}
    \end{equation}
    \item Then there exists a test set $\mathcal{T} = \mathbf{t_{1},..,t_{n}}$ that depends only on the constraint matrix $A$.
    \item This test set can be used to find more optimal points starting from a non-optimal feasible point $\mathbf{x_{o}}$ to the program. This can be done because:
    \begin{equation}\notag
        \exists \; \alpha \in \mathbb{Z_{+}}, \: i\in \{1,..,n\}\; \mid f(\mathbf{x_{o}}+\alpha\mathbf{t_{i}}) < f(\mathbf{x_{o}})
    \end{equation}
     all the while $\mathbf{x_{o}+\alpha \mathbf{t_{i}}}$ still being feasible.
\end{itemize}
\newpage
Now, for what kind of objective functions is the Graver basis a test set? Here is a list of some such objective functions:
\begin{itemize}
    \item Separable convex minimisation:$\sum_{i}f_{i}(\mathbf{c^{T}_{i}x})$ where all the $f_{i}$ are convex functions.
    \item Convex integer minimisation: $-f(\mathbf{Wx}$ where $\mathbf{W}\in \mathbb{Z}^{dxn}$ and $f$ is convex. 
    \item Norm $p$ minimisation: $f(\mathbf{x} = \mid\mid \mathbf{x}-\mathbf{\hat{x}}\mid\mid_{p}$ where $\mathbf{\hat{x}}$ is some vector defined in the domain of $\mathbf{x}$.
    \item Quadratic minimisation: $f(\mathbf{x}=\mathbf{x}^{T}\mathbf{Qx}$ where $\mathbf{Q}$ lies in the dual of the quadratic Graver cone of the constraint matrix $\mathbf{A}$. This also includes certain non-convex $\mathbf{Q}\not\geq0$
    \item Polynomial minimisation:$f(\mathbf{x}) = \textit{P}(\mathbf{x})$ with \textit{P} being a polynomial of degree \textbf{d} that lies on the dual of the $d^{th}$ degree Graver cone of the constraint matrix $\mathbf{A}$.
\end{itemize}
Note that the Graver basis itself does not depend on the objective. The objective just determines whether or not the Graver basis is a test set for the same. This point is beneficial, it allows us to compute a single (possibly non-optimal) feasible point and the Graver basis, and just use these to find an optimal point. This is at the heart of  Graver Augmented Multiseed Algorithm(GAMA - lecture 6).\newline
One other very interesting problem for which the Graver basis is a test set is that of the \textbf{N-fold integer programming}. These are defined as:
\begin{equation}\notag
    \min_{x} \mathbf{wx} \:\mid \: \mathbf{Ax=b},\:l\leq x \leq u, x \in \mathbb{Z}^{n}
\end{equation}
where the constraint matrix $\mathbf{A}$ is of the form:
\begin{equation}\notag
    \mathbf{A} = \mathbf{E}^{(N)} = 
    \underbrace{\begin{pmatrix}
    \mathbf{E}_{1} & \mathbf{E}_{1} & ... & \mathbf{E}_{1} \\
    \mathbf{E}_{2} & 0 & \dots & 0 \\
    0 & \mathbf{E}_{2} & \dots & 0 \\
    \vdots & \vdots & \ddots & \vdots \\
    0 & 0 & \dots & \mathbf{E}_{2}
    \end{pmatrix}}_{N}
\end{equation}
Such problems are widely encountered while dealing with multi-commodity flows, privacy in statistical databases, and closest string determination. But what makes it interesting is that the Graver basis for such a problem can be determined in \textbf{polynomial} complexity! This is unlike the other cases where any current algorithm would still take a worst-case non-polynomial complexity, often exponential.\newline
Using this fact, we could use a similar "lifting" process as in Lawrence lifting and map any constraint matrix to such an $N$-fold matrix. Then we can compute the Graver basis in polynomial time for such a matrix. Of course, though polynomial in time, all this is still pretty computationally expensive as the lifting to the right dimension itself can take a long time. Hence it isn't such a general procedure after all.
\newpage
\section{Graver-based Augmentation}
Based on the idea of using Graver basis as a test set, we arrive at the Graver-based Augmentation algorithm, which is sort of a precursor to GAMA, which will be introduced in lecture 6.
\subsection{Contrasting with the current approaches}
Before jumping into the algorithm, it may be useful to discuss a little about the current classical methods for solving integer programs. These are based on the divide-and-conquer approach, and the one used most popularly is the \textbf{Branch-and-Bound} algorithm. These sorts of algorithms make use of polyhedral theory and give global optimality guarantees. Although they have exponential complexity, there is code available that has been made pretty efficient still. \newline
In contrast, test-set methods rely on "augmentation", wherein we start from a non-optimal feasible point and continually move towards a more optimum point. These make use of the theory of algebraic geometry. They provide a global convergence guarantee and have a polynomial oracle complexity once supplied with a test set.
\subsection{The Algorithm}
The algorithm is as follows:\newline
\begin{addmargin}[1em]{0em}
\textbf{Inputs:}
\begin{itemize}
    \item The constraint matrix $\mathbf{A}$ along with the constraint itself $\mathbf{Ax=b,\; l\leq x\leq u}$.
    \item The test set (Graver basis if applicable), $\mathcal{G}$,  corresponding to the constraint.
    \item The objective function $f$ for the problem.
    \item One feasible point $\mathbf{z}_{o}$ to the problem.
\end{itemize} 
\textbf{Output:} The optimal point $\mathbf{z}_{min}$ to the problem.\newline
\textbf{Loop while}: $\exists \: \mathbf{t}\in \mathcal{G}, \alpha \in \mathcal{Z}_{+} \; \mid \mathbf{z}_{o}+\alpha \mathbf{t}$ is feasible and $f(\mathbf{z}_{o}+\alpha \mathbf{t}) < f(\mathbf{z}_{o})$\newline
\textbf{Do}:\newline
$\alpha, \mathbf{t} = arg\min_{\alpha, \mathbf{t}} f(\mathbf{z}_{o}+\alpha \mathbf{t})$\newline
$\mathbf{z}_{o} = \mathbf{z}_{o}+\alpha \mathbf{t}$\newline
\textbf{done}\newline
\textbf{Return}:$\mathbf{z}_{o}$\newline
\end{addmargin}

This algorithm is demonstrated in the second half of the code \href{https://colab.research.google.com/github/bernalde/QuIP/blob/master/notebooks/Notebook\%203\%20-\%20Graver\%20basis.ipynb}{here}. A walk-through example is also provided in the video lecture corresponding to this lecture.
\newpage
\subsection{Results against Gurobi}
\subsubsection{$N$-fold programming}
Here we showcase some results of using the Graver basis for computing the optimal value for the $N$-fold programs~(closest string program in specific). The results are compared against those from Gurobi:

\begin{figure}[htb]
    \begin{minipage}{.5\textwidth}
        \centering
        \includegraphics[width=0.8\textwidth]{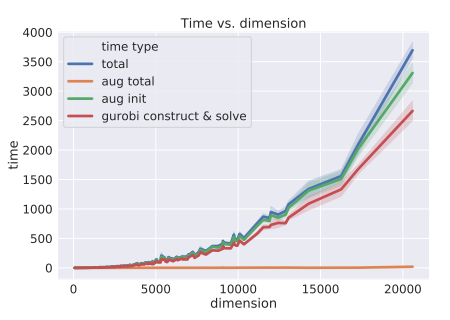}
        \caption{Fig 3.1: Time vs. Size of problem}
        \label{fig:nfold}
    \end{minipage}
    \begin{minipage}{.5\textwidth}
        \centering
        \includegraphics[width=0.8\textwidth]{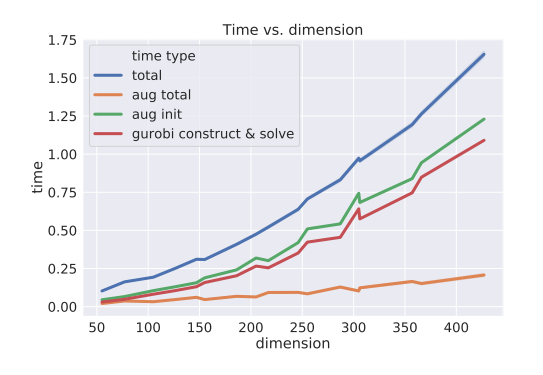}
        \caption{Zoomed in version of figure \ref{fig:nfold}}
        \label{fig:quad}
    \end{minipage}
    \caption{Comparison of Test set augmented methods vs. Gurobi\cite{nfold}}
\end{figure}

A few key points to note here:
\begin{itemize}
    \item aug total -  total augmentation time is the time to calculate the optimal point given the test set. Note that it stays fairly constant, or rather grows slowly.
    \item aug int - The time taken to compute the test set. Notice that this is what takes most of the time in the augmentation algorithm. 
\end{itemize}
Even though we say polynomial time, the computation is still expensive, mostly because of memory issues. If we had a way to compute the Graver basis more efficiently, we would do much better!
\subsubsection{Quadratic Cardinality Boolean}
Here we consider the problem of the type:
\begin{equation}\notag
    \min_{x} \{\mathbf{c^{T}x} + \mathbf{x^{t}Qx} \; \mid \mathbf{1^{T}_{n}x=b}\}
\end{equation}
It turns out, for such a problem statement, the Graver basis can be calculated analytically! Following are some results using the Graver basis, again compared against Gurobi.

\begin{figure}[htb]
    \begin{minipage}{.5\textwidth}
        \centering
        \includegraphics[width=0.8\textwidth]{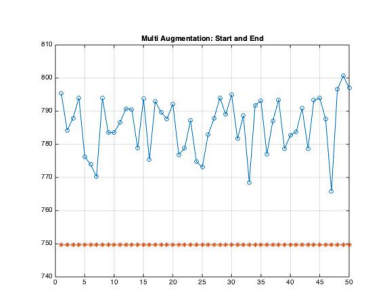}
        \caption{Convex Problem}
        \label{fig:convex}
    \end{minipage}
    \begin{minipage}{.5\textwidth}
        \centering
        \includegraphics[width=0.8\textwidth]{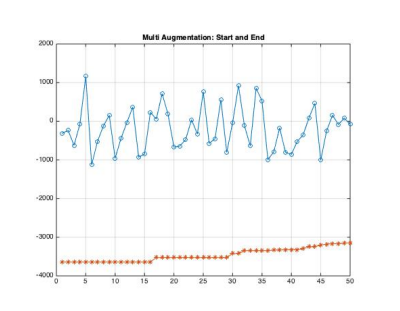}
        \caption{Non-convex Problem}
        \label{fig:non_convex}
    \end{minipage}
    \caption{GAMA convergence from multiple feasible points for convex and non-convex problems\cite{GAMA}}
\end{figure}

Figure~\ref{fig:convex} shows that for convex problems, regardless of what initial feasible point we choose, the optimal is always reached. In figure~\ref{fig:non_convex} we see that for non-convex problems getting to the optimal point is trickier: we land at different points, in the end, depending on our first feasible point. Note that these results are actually for GAMA, but in the first case, we would get the same results for Graver-best augmentation. In the second case, we have to use GAMA to get a good optimum.
\section{Food for Thought}
Some stuff to think about before the next lecture:
\begin{itemize}
    \item We see that computing the test-sets, in general, is classically intractable. Hence just directly applying the same algorithm to every problem instance will not help much. What we could try to do is try tailoring the algorithm to specific problem instances instead.
    \item What if we don't compute the entire Graver basis test set but only a part of it? How good a solution can we reach?
    \item Are there (more) special structured matrices that allow for systematic Graver Basis calculation?
\end{itemize}
Lastly, it worth noting that the Graver basis has applications where current commercial solvers fail, mostly in the case of not-well-behaved objective functions. Hence the potential for it is immense if we find faster ways of solving for the basis or find that, with many feasible solutions, only a partial Graver basis is sufficient.

\lecture{5}{Ising Model}{Prof. Sridhar Tayur, David E. Bernal}{Sashank Kaushik Sridhar}
{https://bernalde.github.io/QuIP/slides/47-779\%20Lecture\%204\%20-\%20Ising\%20Model.pdf}
{https://colab.research.google.com/github/bernalde/QuIP/blob/master/notebooks/Notebook\%204\%20-\%20Ising\%20Model.ipynb}
{https://cmu.zoom.us/rec/share/AOmRlMOFoAuTtyuyEDRQRMu1LuhHZJiY9AjUc3CZ-XMyP3nc3ZxJmop90Hn7wdI.R-a68jRiL_Ov3UPE}
{jKYs*s8@}

\section{Introduction \& Background}
The Ising model was first introduced by Lenz (1920) to explain the ferromagnetic-paramagnetic phase transition. It was solved by his student Ising (1925) for the 1-D case, followed by Onsager (1940) for the 2-D case. The 3-D case was reformulated as a \emph{non-planar graph MAXCUT problem}, proving the 3-D case to be \emph{NP-Complete}. By universality of dimensions and symmetries, Ising models have been shown to have important implications for non-magnetic phase transitions as well.

\begin{figure}
    \centering
    \includegraphics[scale = 0.90]{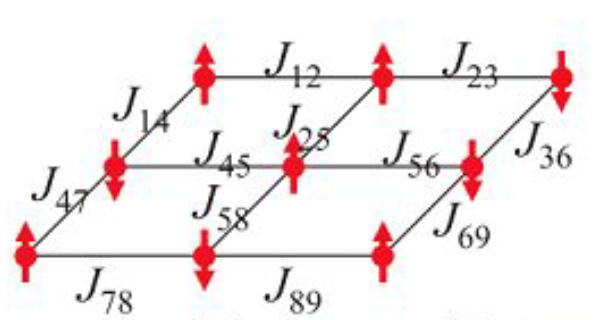}
    \caption{Typical 2-D Ising Lattice.}
    \label{fig:2d_ising}
\end{figure}

\subsection{Mathematical Definition}
\[
H(\bm{\sigma}) = -\sum_{(ij)\in E(G)}J_{ij}\sigma_i\sigma_j - \mu\sum_{i\in V(G)}h_i\sigma_i
\]
where,
\begin{itemize}
    \item $H$: Hamiltonian
    \item $\bm{\sigma} \in \{-1,+1\}^{V(G)}$: Spin per lattice site
    \item $G = (V,E)$: Graph or Lattice defining the interactions
    \item $\mu$: Magnetic Moment
    \item $J_{ij}$: Interaction between sites
    \item $h_i$: External field applied per site
\end{itemize}
Note:- Depending on the signs of $J_{ij}$ and $h_i$, the interactions can be ferromagnetic or antiferromagnetic.
The partition function gives the probability of each configuration ($Z(\beta)$),
\[
P(\bm{\sigma},\beta) = e^{-\beta H(\bm{\sigma})}/ Z(\beta) 
\]
where,
\begin{itemize}
    \item $\beta = (k_BT)^{-1}$
    \item $Z(\beta) = \sum_{\bm{\sigma}} e^{-\beta H(\bm{\sigma})}$
\end{itemize}

\subsection{Solutions to Ising Models}
Solutions exist in 1-D for both circular and free-boundary conditions without external fields. However, in such solitions
np phase transition is observed. With the external field applied, the phase transition is observed when $J=h$. In the 2-D case, Onsager's solution predicts a phase transition. In 3-D, the non-planar graph is NP-Complete via the MAXCUT formulation, but mean-field approximations exist. Different approaches can be adapted depending on the problem's objectives, be it finding important statistical properties or the exact spin configurations.

\section{Reformulations}
As stated, the Ising model can be recast as a MAXCUT problem, a QUBO, and thereby as an Integer Linear Program.
\\
For the MAXCUT reformulation, we start with no external field,\\
\begin{displaymath}
    \begin{aligned}
        H(\bm{\sigma}) & = -\sum_{(ij)\in E(G)}J_{ij}\sigma_i\sigma_j\\&
        = -\sum _{(ij)\in E(V^{+})}J_{ij}-\sum _{(ij)\in E(V^{-})}J_{ij}+\sum _{(ij)\in \delta (V^{+})}J_{ij}\\&
        = -\sum _{(ij)\in E(G)}J_{ij}+2\sum _{(ij)\in \delta (V^{+})}J_{ij}\\&
        = C+2\sum _{(ij)\in \delta (V^{+})}J_{ij}
    \end{aligned}
\end{displaymath}

Taking $W_{ij} = -J_{ij}$ as the weights of the edges, we recast the minimization of the Hamiltonian as a maximum cut of the graph.\\
Integer Linear Program:\\
\begin{displaymath}
    \begin{aligned}
        min_{x\in \{0,1\}^n} &\sum_{(ij)\in E(G)}Q_{ij}x_{ij} + \sum_{i\in V(G)}Q_{ii}x_i + c\\
        s.t.\;\;\;& x_{ij} \geq x_i+x_j-1\\
        & x_{ij}\leq x_i\\
        & x_{ij}\leq x_j\\
        & x_i,x_{ij}\in \{0,1\}\; \forall \;(ij) \in E(G), i \in V(G)\\
    \end{aligned}
\end{displaymath}
with $Q_{ij}=4J_{ij}, Q_{ii}= 2h_i - \sum_{j\in V(G)}(2J_{ij}+2J_{ji}), c=\sum_{i<j}J_{ij}-\sum_{i\in V(G)}h_i$

\section{Solving Ising Models}
Ising models can be solved computationally using the Markov-Chain Monte Carlo Metropolis-Hastings Algorithm which we describe next.

\subsection{Metropolis-Hastings Monte Carlo~(MHMC) Algorithm}
\begin{enumerate}
    \item Start with a known spin configuration $\bm{\sigma}^i$, with corresponding energy $H(\bm{\sigma}^i)$ and temperature $T = (k_B\beta)^{-1}$.
    \item Randomly change spin configuration: $\bm{\sigma}^j = \bm{\sigma}^i + \bm{\delta}$
    \item Calculate new $H(\bm{\sigma}^j)$.
    \item Compare to the previous energy and,
    \begin{itemize}
        \item[-] If $H(\bm{\sigma}^j)< H(\bm{\sigma}^i)$, keep new position
        \item[-] If $H(\bm{\sigma}^j)> H(\bm{\sigma}^i)$, keep new position if $\mathrm{exp}\left[-\frac{H(\bm{\sigma}^j)- H(\bm{\sigma}^i)}{k_BT}\right] \geq \mathrm{Rand}[0,1]$.
    \end{itemize}
    \item Repeat steps 2-4 \textit{K} times.
\end{enumerate}

\subsection{Simulated Annealing}
The MHMC algorithm can be improved by providing a heuristic to lower the temperatures in a scheduled fashion. This algorithm
is based on a metallurgical process of annealing, hence the name \emph{simulated annealing}. If the temperature is lowered slowly enough, the algorithm is guaranteed to reach the ground state of the energy profile.
The default limits for the temperature are given by $\beta \in \left[ \frac{\ln(2)}{\max \{ \Delta E \} },\frac{\ln(100)}{\min \{ \Delta E \} } \right]$.
At every temperature, we run the MHMC algorithm as a nested loop. The outer loop incrementally reduces the temperature from an effectively high starting point.
\begin{figure}
    \centering
    \includegraphics[scale = 0.77]{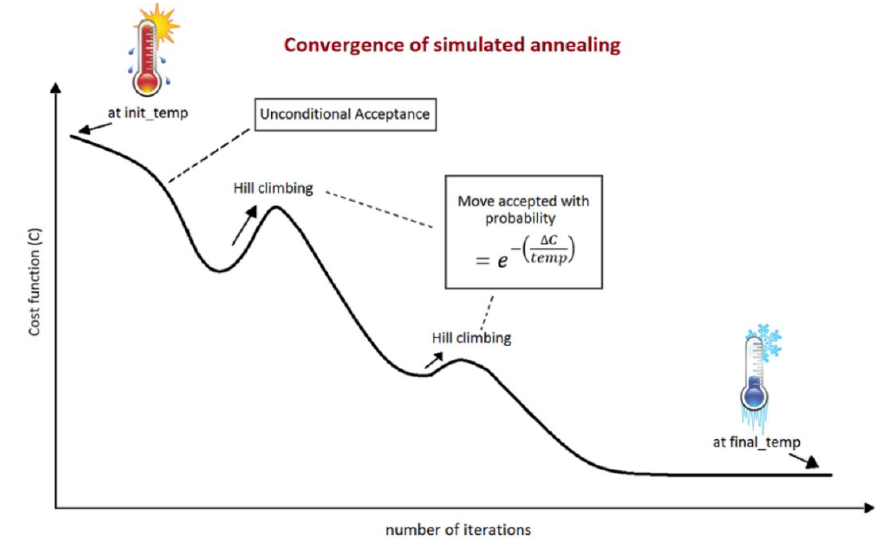}
    \caption{Traversing the energy landscape through simulated annealing.}
    \label{fig:sim_ann_landscape}
\end{figure}
\subsection{Advanced Simulated Annealing}
For large graphs, the update $\delta$ is a limiting factor in the algorithm's runtime. If individual spins are updated, the algorithm can take an exponential time to converge to the global optimum. Hence, we update the spins in clusters.
\begin{itemize}
    \item This must be done carefully not to violate energy conservation and ensure that the evolution is ergodic. Any state can be reached from another via the Markov-Chain.
    \item To ensure convergence, we generate multiple replicas of the system at different temperatures. After a fixed number of MHMC updates, the temperatures of two replicas $r_1,r_2$ are exchanged with a probability:\\
    $P(r_1 \leftrightarrow r_2) = \mathrm{min}\{1, \mathrm{exp}[(\beta_1-\beta_2)(H(\bm{\sigma}^1)- H(\bm{\sigma}^2))]\}$
    \item Two temperatures are always exchanged if a replica at higher temperature has a lower energy than a replica with a lower temperature.
    \item Otherwise, the exchange of the two temperatures is either accepted or rejected using the random number between 0 and 1
\end{itemize}

\section{Comparing Heuristics}
Comparing different heuristics is generally difficult due to several parameters to tune and differences in the underlying hardware. Two metrics to consider are Time and Solution Quality. A useful metric is given as Time To Solution (TTS):
\[
\text{TTS}(m)= m\tau(m)\frac{\log(1-s)}{\log(1-p(m))}
\]

\begin{itemize}
    \item $m$: Number of times run, or sweeps in Simulated Annealing
    \item $s$: Success probability after $m$ sweeps
    \item $p(m)$: Probability of success to achieve (usually high)
    \item $\tau(m)$: Time it takes to perform a single sweep
\end{itemize}

Several heuristics exist for MAXCUT and QUBO. These heuristics can be compared with each other using this metric. Simulated Annealing algorithms, as highlighted below, seem to perform well among other heuristics.
\begin{figure}
    \centering
    \includegraphics[scale = 0.70]{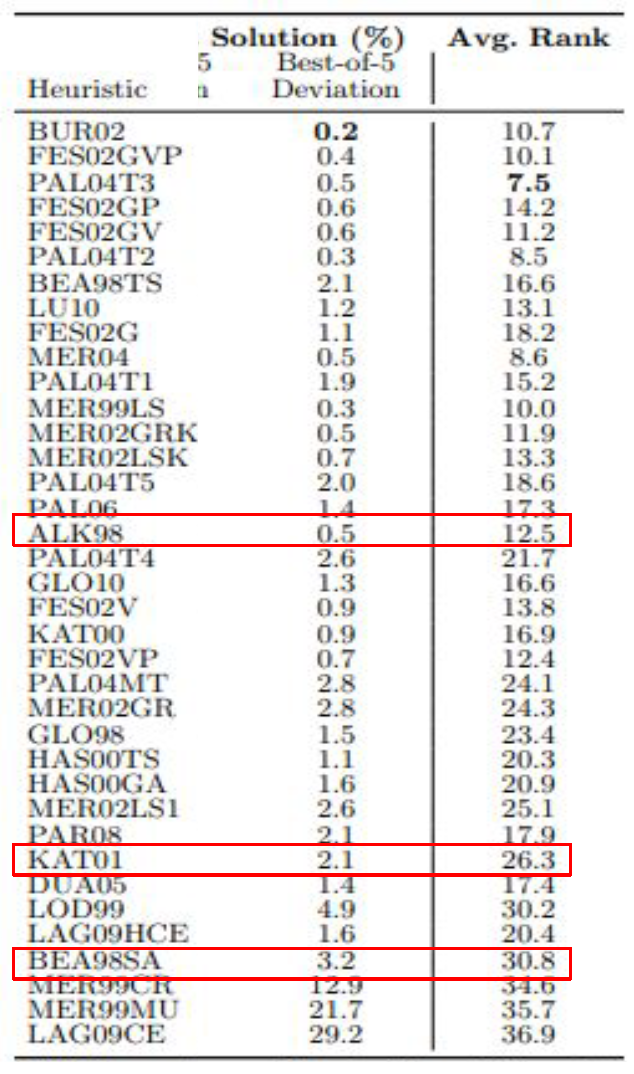}
    \caption{Heuristic methods for addressing QUBO problems, with the ones using Simulated Annealing highlighted \cite{dunning2018works}}
    \label{fig:heuristics_sim_ann}
\end{figure}

\lecture{6}{QUBO - Quantum Unconstrained Binary Optimization}{Prof. Sridhar Tayur, David E. Bernal}{Suriyaa Valliapan, EP17B027}
{https://bernalde.github.io/QuIP/slides/47-779\%20Lecture\%205\%20-\%20Quadratic\%20Unconstrained\%20Binary\%20Optimization\%20(QUBO).pdf}
{https://colab.research.google.com/github/bernalde/QuIP/blob/master/notebooks/Notebook\%205\%20-\%20QUBO.ipynb}
{https://youtu.be/v3D5jxi_2wM}
{}




\section{Abstract}
This lecture introduces us to QUBO and its equivalence to the Ising model. We then move on to reformulating a general Integer Programming problem as a QUBO problem, ending with a demonstration.

\section{QUBO Model}
QUBO expands to Quadratic Unconstrained Binary Optimization. Quadratic, because it contains expressions of order 2; unconstrained, because there are no constraints; and binary, since all variables are either zero or one. Equation (1) represents the QUBO model. 
\begin{equation}
\begin{array}{l}
\min _{\mathbf{x} \in\{0,1\}^{n}} \sum_{(i j) \in E(G)} x_{i} Q_{i j} x_{j}+\sum_{i \in V(G)} Q_{i i} x_{i}+c \\
=\min _{\mathbf{x} \in\{0,1\}^{n}} \mathbf{x}^{\top} \mathbf{Q} \mathbf{x}+c
\end{array}
\end{equation}
The offset $c$ above is irrelevant for the optimization since it is independent of the variables. The problem above represents an integer program that is non-linear (although it could be linearised). If $\mathbf{Q}$ is not positive semi-definite, the problem may be non-convex.

\subsection{Quadratic Coefficient Matrix Q}
The \textbf{$Q_{i_j}$} in equation (1) represents the Quadratic Coefficient Matrix. There is no unique representation for this matrix. The matrix \textbf{$Q$} can be upper triangular or complete because of the fact that $x_i x_j= x_j x_i$. We can also note that the elements on the diagonal can be linearised since $x_i ^2 = x_i$ if $x \in \{0,1\}$. This matrix can also be interpreted as the adjacency matrix of a graph (weighted edge between node $x_i$ and $x_j$).

\section{QUBO Ising mapping}
The Ising problem can be converted to a QUBO as follows: Transform the variables $x_i \in \{0,1\}$ to $ \sigma_i \in \{-1,1\}$ such that 
\begin{equation}
\begin{array}{l}
\sigma_{i} =2 x_{i}-1, \\
\sigma_{i} \sigma_{j}=4 x_{i} x_{j}-2 x_{i}-x_{j}+1.
\end{array}
\end{equation}
In these new variables, the objective function becomes
\begin{equation}
\begin{array}{c}
\min _{\sigma \in\{-1,+1\}^{n}} \sum_{(i j) \in E(G)} J_{i j} \sigma_{i} \sigma_{j}+\sum_{i \in V(G)} h_{i} \sigma_{i}+c_{I}= \\
\min _{\mathbf{x} \in\{0,1\}^{n}} \sum_{(i j) \in E(G)} x_{i} Q_{i j} x_{j}+\sum_{i \in V(G)} Q_{i i} x_{i}+c_{Q} \\ \\
Q_{i j}=4 J_{i j}, Q_{i i}=2 h_{i}-\sum_{j \in V(G)}\left(2 J_{i j}+2 J_{j i}\right), c_{I}=c_{Q}+\sum_{i<j} J_{i j}-\sum_{i \in V(G)} h_{i}
\end{array}
\end{equation}
where E(G) is the set of pairs of nodes on the graph G that are connected (Edges), and V(G) is the set of vertices of the graph G. Similarly, a QUBO problem can be converted to an Ising problem as follows:
\begin{equation}
\begin{array}{l}
x_{i}=\left(\sigma_{i}+1\right) / 2 \\
x_{i} x_{j}=\left(\sigma_{i} \sigma_{j}+\sigma_{i}+\sigma_{j}+1\right) / 4
\end{array}
\end{equation}

Once we have this variable transformation, the objective function becomes,

\begin{equation}
\begin{array}{c}
\min _{\mathbf{x} \in\{0,1\}^{n}} \sum_{(i j) \in E(G)} x_{i} Q_{i j} x_{j}+\sum_{i \in V(G)} Q_{i i} x_{i}+c_{Q}= \\
\min _{\sigma \in\{-1,+1\}^{n}} \sum_{(i j) \in E(G)} J_{i j} \sigma_{i} \sigma_{j}+\sum_{i \in V(G)} h_{i} \sigma_{i}+c_{I} \\ \\
J_{i j}=Q_{i j} / 4, h_{i}=Q_{i i} / 2+\sum_{j \in V(G)}\left(Q_{i j} / 4+Q_{j i} / 4\right), c_{I}=c_{Q}+\sum_{i<j} Q_{i j} / 4-\sum_{i \in V(G)} Q_{i i} / 2
\end{array}
\end{equation}
Thus, we have shown that the QUBO and Ising are equivalent models. 

\section{QUBO-Integer Programming Model}
Our goal is to formulate a generic integer programming problem as a QUBO. Here, for the sake of completeness, we talk about QUBO as an Integer programming model. 
\begin{equation}
\min _{\mathbf{x} \in\{0,1\}^{n}} \mathbf{x}^{\top} \mathbf{Q} \mathbf{x}+c
\end{equation}
This non-integer program is solvable using commercial INLP solvers such as Gurobi. To simplify things, we convert this to an ILP by adding new variables $x_{ij} = x_i x_j$. The non-linearity in these can be captured by using linear inequalities as shown below. Experimental results show that this is the most efficient ILP formulation of QUBO.\\
\begin{equation}
\begin{array}{l}
\min _{\mathbf{x} \in\{0,1\}^{n}} \sum_{(i j) \in E(G)} Q_{i j} x_{i j}+\sum_{i \in V(G)} Q_{i i} x_{i}+c \\ \\
\text { s.t. } x_{i j} \geq x_{i}+x_{j}-1, x_{i j} \leq x_{i}, x_{i j} \leq x_{j} \quad \forall(i j) \in E(G) \\
x_{i} \in\{0,1\} \quad \forall i \in V(G), x_{i j} \in\{0,1\} \quad \forall(i j) \in E(G)
\end{array}
\end{equation}
\section{General Integer Programming Problem as a QUBO}
A generic integer programming problem can be converted to a QUBO model as follows. We convert all the integer variables to binary variables, a process called Binarization. All  non-linearities are converted to quadratic terms, a process
called quadratization, and finally one performs  unconstraining, i.e, removing all the constraints by introducing penalties in the objective function. 
\subsection{Binarization}
Our goal is to transform all integer variables to binary variables. Let us denote a single positive integer variable by $y$ which can take values in the range ${0,1,.....y'}$. Now, $y$ can be written in terms of binary variables $x$ as,
\begin{equation}
y=\sum_{j=1}^{d} k_{j} x_{j}=\mathbf{k}^{\top} \mathbf{x}, k_{j} \in \mathbb{Z}_{+}, x_{j} \in\{0,1\}
\end{equation}
with $d$ being the width of the integer encoding. The most obvious choice for expressing an integer variable in terms of binary variable is the \textbf{binary encoding} where $k_{j}=2^{j-1}, d=\left\lfloor\log _{2}(\bar{y})\right\rfloor$. The main issue with this is that the coefficients become very huge which our quantum annealer may not be able to handle with a good precision. 

The other choice is to use an \textbf{unary encoding} where we set all the $k_j = 1$ and the width $d$ becomes $y'$. The problem with this is that the width becomes very large, again causing issues. 

The choice which fits best here is the \textbf{bounded encoding} where we find an encoding with an upper bound for the coefficients ($\mu << y'$). We will not go into further details of this although we just state it here for the sake of completion. \begin{equation}
\begin{array}{c}
\text { if } \bar{y}<2^{\lfloor\log (\mu)\rfloor}+1 \\
\mathbf{k}=\left[2^{0}, 2^{1}, \ldots, 2^{[\log (\bar{y})]-1}, \bar{y}-\sum_{i=1}^{[\log (\bar{y})]} 2^{i-1}\right]
\end{array}
\end{equation}

and, 

\begin{equation}
\begin{array}{c}
\text { if } \bar{y}>2^{\lfloor\log (\mu)\rfloor}+1 \\
\rho=\lfloor\log \mu\rfloor+1, v=\bar{y}-\sum_{i=1}^{\rho} 2^{i-1}, \text { and } \eta=\left\lfloor\frac{v}{\mu}\right\rfloor
\end{array}
\end{equation}

\begin{equation}
k_{i}=\left\{\begin{array}{ll}
2^{i-1} & \text { for } i=1, \ldots, \rho \\
\mu & \text { for } i=\rho+1, \ldots, \rho+\eta \\
v-\eta \mu & \text { for } i=\rho+\eta+1 \text { if } v-\eta \mu \neq 0
\end{array}\right.
\end{equation}

Noise is a significant issue in quantum annealers. Karimi et al., in their paper on 'Practical Integer-to-Binary mapping for quantum annealers', showed that bounded encoding gave the best performance, as can be seen from figures (1) and (2). The probability of getting the right solution decreases exponentially with the noise's standard deviation for the binary case, as seen in figure (1). In contrast, figure (2) shows that the unary encoding takes much time for reaching the solution. It grows with increase in the standard deviation of the noise. Hence, we can see that bounded encoding is the better of the three.

\begin{figure}[h!]
\centering
\includegraphics[width=70mm]{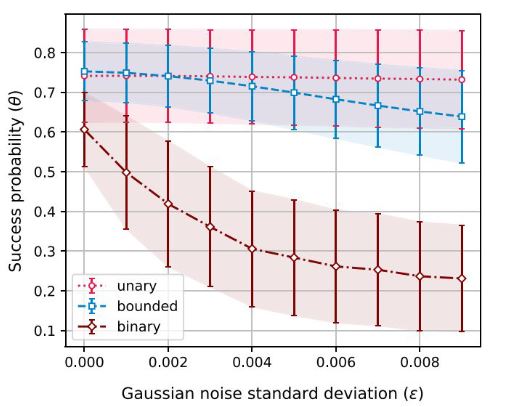}
\caption{The scaling of success probability as a function of standard deviation of noise}
\label{fig:method}
\end{figure}

\begin{figure}[h!]
\centering
\includegraphics[width=70mm]{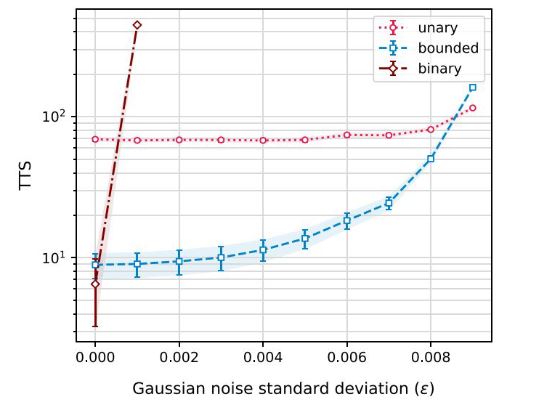}
\caption{The scaling of time to solution as a function of standard deviation of noise}
\label{fig:method}
\end{figure}

\subsection{Quadratization}
We state a theorem which is essential for our further discussion. 

\subsubsection{Theorem}
Any pseudo-Boolean function defined as $f:\{0,1\}^n \to \mathbb{R}$ can be written uniquely as a sum of multi-linear functions. 

\begin{equation}
f(x)=a_{0}+\sum_{i} a_{i} x_{i}+\sum_{i j} a_{i j} x_{i} x_{j}+\sum_{i j k} a_{i j k} x_{i} x_{j} x_{k}+\ldots
\end{equation}

We define new variables $x_{ij}$. We do that for all terms of order greater than 2 in the above function continuously until we are left with just quadratic terms. Hence, by using the above theorem and this observation, we can conclude that any such objective function can be converted to a quadratic polynomial. In an earlier section, we had defined the $x_{ij}$ as $x_i x_j$ and imposed some additional constraints to capture the non-linearity. Instead, here we define a function $H$:

\begin{equation}
    H(\textbf{x}) = 3 x_{ij} + x_i x_j - 2 x_{ji} x_i - 2 x_{ij} x_j   
\end{equation}
which satisfies $H(\textbf{x}) = 0$ if and only if $x_{ij} = x_i x_j$. Hence we can add $H(\textbf{x}) = 0$ to our constraints instead of the three inequalities, thus simplifying our problem further. The variables $x_{ij}$ are called ancillary variables. We could also represent these new constraints through graphs.

\begin{figure}[h!]
\centering
\includegraphics[width=30mm]{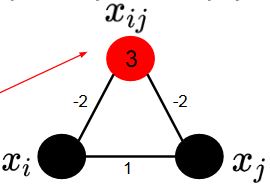}
\caption{Representation of the constraint $H(\textbf{x}) = 0$ through a graph. Here the edges denote the coefficient of the product of the variables it connects in the constraint function. The value inside the red circle represents the coefficient of just the variable corresponding to that node.}
\label{fig:method}
\end{figure}

We had written $H(\textbf{x})$ for the case when we had terms of the form $x_i x_j$ which is equivalent to logical AND. We can come up with similar expressions for OR and NOT gates, and once we have the constraints for these, we could come with expressions for everything else. This idea can also be easily extended to higher orders, as shown in figure (4).

\begin{figure}[h!]
\centering
\includegraphics[width=120mm]{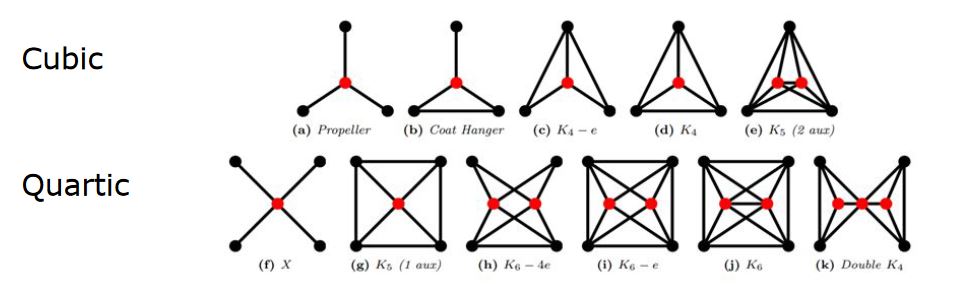}
\caption{Representation of constraints for higher order problems using graphs}
\label{fig:method}
\end{figure}

\subsection{Unconstraining}
Previously, we had removed the constraints in our linear programming problem using the concept of Lagrange multipliers. Using that concept, we moved our constraints into the objective function by weighting/ penalizing them with Lagrange multipliers. To demonstrate this, 
\begin{equation}
\begin{array} {c} 
min_{\textbf{x} \geq 0} \ \textbf{c}^T \textbf{x} 
\\
\text{s.t} \  \textbf{Ax} = \textbf{b}
\end{array}
\end{equation}
can be written as,
\begin{equation}
min_{\textbf{x} \geq 0} \ \textbf{c}^T \textbf{x} + \lambda^T (\textbf{b} - \textbf{Ax})
\end{equation}

As explained in the previous lectures, we can turn this optimization problem into a problem of finding these Lagrange multipliers. For linear programs, we had this very nice principle of strong duality. This principle states that the optimal objective of our initial~(primal) problem is the same as the optimal objective of the dual problem,
\begin{equation}
\begin{array}{c}
max_{\lambda} \ \mathbb{L} = max_{\lambda} \ \lambda^T \textbf{b} 
\\
\text{s.t} \ \lambda^T \textbf{A} \leq \textbf{c}^T
\end{array}
\end{equation}
In other words,
\begin{equation}
\lambda^{\ast T} \textbf{b} = \textbf{c}^T \textbf{x}
\end{equation}

However for integer programs, strong duality doesn't hold. However, we can still use the multipliers to convert the problem into an unconstrained one, as demonstrated in the following example. In this example, we move constraints into the objective~(note the signs of the multipliers)
\begin{equation}
\begin{array} {c} 
min_{\textbf{x} \in \{0,1\}^n} \ \textbf{c}^T \textbf{x} 
\\
\text{s.t} \  g(\textbf{x}) \leq 0 \ (\lambda)
\\
h(\textbf{x}) = 0 \ (\rho)
\end{array}
\end{equation}
can be written as,
\begin{equation}
\begin{array}{c}
min_{\textbf{x} \in \{0,1\}^n} \ \textbf{c}^T \textbf{x} + \lambda^T g(\textbf{x}) + \rho^T h(\textbf{x}) \\
\text{where} \ \lambda_i^T \leq 0
\end{array}
\end{equation}
\subsection{Putting everything together}
Now, using all the tools explained above, we demonstrate how to reformulate an integer programming model as a QUBO. \\
\begin{itemize}
    \item Step 1: Binarize \\
    \begin{equation}
    \begin{array}{c}
    m \min_{\textbf{x} \in \{0,1\}^n} \, \textbf{c}^T \textbf{x} 
    \\
    \text{s.t} \,  g(\textbf{x}) \leq 0  (\lambda)
    \\
    h(\textbf{x}) = 0 \, (\rho)
    \end{array}
    \end{equation}
    
    \item Step 2: Quadratize (getting new variables y)
    \begin{equation}
\begin{array}{c}
\min _{\mathbf{y} \in\{0,1\}^{n}} \mathbf{d}^T \mathbf{y} \\
\text { s.t. } \mathbf{y}^{\top} G \mathbf{y} \leq 0 \\
\mathbf{y}^{\top} H \mathbf{y}=0
\end{array}
\end{equation}

    \item Step 3: Transform inequalities into equalities
    \\ Introduce slack variables $s$ to convert the inequalities to equalities so that we get, (Note that by scaling the constraints, the slack variables become integers)
    \begin{equation}
\begin{array}{c}
\min _{\mathbf{y} \in\{0,1\}^{n}, \mathbf{s} \in \mathbb{Z}_{+}} \mathbf{d}^{\top} \mathbf{y} \\
\text { s.t. } \mathbf{y}^{\top} G \mathbf{y}+\mathbf{s}=0 \\
\mathbf{y}^{\top} H \mathbf{y}=0 \quad(\rho)
\end{array}
\end{equation}

    \item Step 4: Unconstraining
    \\ Considering that we can only deal with discrete variable in QUBO, we cannot change the value of the multipliers. So, we set the multipliers to be constant penalty factors. In our case, Integer linear equalities become penalties by squaring the constraints (We square them to avoid the case where the minimisation of the term leads to negative infinities). So, we get,
\begin{equation}
\min _{\mathbf{y} \in \mathbb{Z}_{+}, \mathbf{s} \in \mathbb{Z}_{+}} \mathbf{d}^{\top} \mathbf{y}+\rho (\mathbf{y}^{\top} G \mathbf{y}+\mathbf{s})^{\top}(\mathbf{y}^{\top} G \mathbf{y}+\mathbf{s}) + \lambda (\mathbf{y}^{\top} H \mathbf{y})^{\top}(\mathbf{y}^{\top} H \mathbf{y})
\end{equation}
\end{itemize}
\subsection{Example}
Example optimization problem:
\begin{equation}
\begin{array} {c} 
min_{\textbf{x} \geq 0} \ \textbf{c}^T \textbf{x} 
\\
\text{s.t} \  \textbf{Ax} = \textbf{b}
\end{array}
\end{equation}
Let us represent the constraint as $H_A = 0$ and the objective function by $H_B$
\begin{equation}
H_{A}=\rho \sum_{j=1}^{m}\left(\sum_{i=1}^{n} A_{i j} x_{i}-b_{j}\right)^{2}
\end{equation}
\begin{equation}
H_{B}=\sum_{i=1}^{n} c_{i} x_{i}
\end{equation}
Hence, the total energy function of the QUBO can be written as, $H = H_A + H_B$. The next question we encounter is how to determine the penalty factor $\rho$. These factors reflect the weightage given to the constraints. We want it to be  such that any infeasible  solution that has the lowest value of $H_B$ should still have an overall higher value of $H$ than the best feasible solution (satisfies the constraints and has the lowest value of $H_B$). We express this quantitatively as follows: we want even a reasonably small enough infeasibility to be heavily penalized, this is required because our quantum annealers are noisy they have a limited precision and a limited range of fidelity
\begin{equation}
    \rho \geq \frac{\Delta H_B^{max}}{\Delta H_A^{min}}
\end{equation}
where,
\begin{equation}
\begin{array}{l}
\Delta H_{B}^{\max }=\sum_{i=1}^{n} \max \left\{c_{i}, 0\right\} \\ \\
\Delta H_{A}^{\min }=\min _{\sigma_{i} \in\{0,1\}, j}\left(\max \left[1, \frac{1}{2} \sum_{i=1}^{n}(-1)^{\sigma_{i}} A_{i j}\right]\right)
\end{array}
\end{equation}
\section{Demonstration}
In the \href{https://colab.research.google.com/github/bernalde/QuIP/blob/master/notebooks/Notebook%205%20-%20QUBO.ipynb#scrollTo=DtLBNn2Osq9j}{Notebook} a couple of simple problems formulated as QUBO are solved using simulated annealing. 
\begin{itemize} 
    \item Problem 1 \\
    \begin{equation}
        \begin{array}{c}
    \min_{\mathbf{x}} 2 x_0+4 x_1+4 x_2+4 x_3+4 x_4+4 x_5+5 x_6+4 x_7+5 x_8+6 x_9+5 x_{10} \\
s.t. 
\begin{bmatrix}
1 & 0 & 0 & 1 & 1 & 1 & 0 & 1 & 1 & 1  \\
0 & 1 & 0 & 1 & 0 & 1 & 1 & 0 & 1 & 1  \\
0 & 0 & 1 & 0 & 1 & 0 & 1 & 1 & 1 & 1 
\end{bmatrix} \mathbf{x}=
\begin{bmatrix}
1\\
1\\
1
\end{bmatrix} \\
\mathbf{x} \in \{0,1 \}^{11}
\end{array}
\end{equation}
In order to define the $Q$ matrix, we write the problem
\begin{equation}
\begin{array}{c}
\min_{\mathbf{x}} \mathbf{c}^\top \mathbf{x}\\
s.t. \mathbf{A}\mathbf{x}=\mathbf{b} \\
\mathbf{x} \in \{0,1 \}^{11}
\end{array}
\end{equation}
as follows:
\begin{equation}
\begin{array}{c}
\min_{\mathbf{x}} \mathbf{c}^\top \mathbf{x} + \rho(\mathbf{A}\mathbf{x}-\mathbf{b})^\top (\mathbf{A}\mathbf{x}-\mathbf{b}) \\
\mathbf{x} \in \{0,1 \}^{11}
\end{array}
\end{equation}
Exploiting the fact that $x^2=x$ for $x \in \{0,1\}$ we absorb linear terms in the diagonal of the $Q$ matrix.
\begin{equation}
\begin{array}{c}
\rho(\mathbf{A}\mathbf{x}-\mathbf{b})^\top (\mathbf{A}\mathbf{x}-\mathbf{b}) = \rho( \mathbf{x}^\top (\mathbf{A}^\top \mathbf{A}) \mathbf{x} - 2\mathbf{b}^\top \mathbf{A} \mathbf{x} + \mathbf{b}^\top \mathbf{b} )
\end{array}
\end{equation}
For this problem in particular, one can prove the the penalization factor is given by $\rho > \sum_{i=1}^n |c_i|$, therefore we choose this bound + 1.
\\
Remarks:
\begin{itemize}
    \item The annealer gets stuck in a local minima majority of the time.
    \item The constraints may not be strictly satisfied as we have just added these terms as penalties to the objective function. The whole expression being minimized does not necessarily mean that the constraints have to be satisfied. For constraints to be satisfied exactly, we have to invoke tools from algebraic geometry (G{"o}ebner basis). Consider the quadratic polynomial where the goal is to solve (obtain its zeros) $H_{i j}$  as a QUBO. This technique will allow us to eliminate slack variables. 
    \begin{equation}
    H_{i j}:=Q_{i} P_{j}+S_{i, j}+Z_{i, j}-S_{i+1, j-1}-2 Z_{i, j+1}
    \end{equation}
    Now, compute the Gr{"o}ebner basis of the system, 
    \begin{equation}
\mathcal{S}=\left\{H_{i j}\right\} \cup\left\{x^{2}-x, x \in\left\{P_{j}, Q_{i}, S_{i, j}, S_{i+1, j-1}, Z_{i, j}, Z_{i, j+1}\right\}\right\}
\end{equation}
\end{itemize}
and look for a positive quadratic polynomial. 
\begin{equation}
H_{i j}^{+}=\sum_{t \in \mathcal{B} \mid \operatorname{deq}(t)<2} a_{t} t
\end{equation}
Global minima of $H_{i j}^+$ are the zeros of $H_{i j}$. (Solve for $a_t$)
\item Problem 2  \\ 
$\mathcal{G}(V, E)$ has 12 vertices and 23 edges. We ask if the graph is $3$–colorable. (Graph Coloring problem)
Remarks:
\begin{itemize}
    \item Because of precision issues in the translation to QUBO, we obtain very tiny coefficients that should be zero. In any case, since this is a constraint satisfaction problem, any of the solutions with energy $~0$ is a valid coloring.
\end{itemize}
\end{itemize}
\section{Summary}
\begin{itemize}
    \item Introduction to QUBO model and a brief description of the model. 
    \item A QUBO model is equivalent to the Ising problem.
    \item QUBO problem is an INLP, and it may or may not be convex.
    \item A general Integer Programming Problem can be reformulated as a QUBO by binarization of the variables, Quadratization of the Objective and the constraints, converting the inequalities to equalities using slack variables, and then unsconstraining by adding the constraints as penalties to the Objective function. 
    \item A couple of examples and the limitations, difficulties, and the subtleties involved. 
\end{itemize}

    
    
    
    
    
    
    
\lecture{7}{GAMA - Graver Augmented Multiseed Algorithm}{Prof. Sridhar Tayur, David E. Bernal}{Gautham Umasankar}
{https://bernalde.github.io/QuIP/slides/47-779\%20Lecture\%206\%20-\%20Graver\%20Augmented\%20Multiseed\%20Algorithm\%20(GAMA).pdf}
{https://colab.research.google.com/github/bernalde/QuIP/blob/master/notebooks/Notebook\%206\%20-\%20GAMA.ipynb}
{https://cmu.zoom.us/rec/share/_oLw25EljDXFHdjcg3LQdHD09pdyNDQua8kKv9vEUnlxKbRk1XmsFXdGegiajaky.cfnaZLsRitlNKoNL}
{3\&UUXKgf}



\section{Introduction - Hybrid Quantum-Classical Algorithms}

A naive way of solving an integer program given by:
\begin{align}
    &min\: f(x) \\ 
    &Ax = b 
   \\ &l\leq x \leq u
   \\ &x\in \mathbb{Z}^{n}
\end{align}
is to first quadratize $f(x)$ as $x^{T}Qx$ and add the constraint using a Lagrange multiplier to get:
\begin{equation}
    min\: x^{T}Qx + \lambda(Ax - b)^{T}(Ax-b).
\end{equation}
However, this has balancing issues and makes the objective even more complicated by increasing its size. We want to take advantage of the annealer, but more intelligently.  

\subsection{GAMA: A Hybrid Quantum-Classical Optimization Algorithm}

GAMA is an algorithm where the classically expensive Graver Basis computation is done by Ising solvers like the D-Wave Quantum Annealer, Simulated Bifurcation Machine, etc. The Ising solver is also used to find many initial feasible solutions, which are then augmented using the Graver Basis to find the optimal solutions. \\  
The defining feature of this algorithm is that it uses the objective function just like an oracle call. The objective function doesn't enter the QUBO formulation at all. This allows GAMA to be applied to complex objective functions. 
A broad outline of the steps involved in GAMA: \cite{quantum_GAMA}

\begin{enumerate}
    \item Calculate the Graver Basis (Quantum, with Classical post-processing)
    \item Find many initial feasible solutions or "seeds" (Quantum)
    \item Augment each feasible solution using the Graver Basis computed in step 3 (Classical)
\end{enumerate}

\section{Test sets in Optimization, a recap}
Given an integer program:
\begin{align}
    &min\: f(x) \\ 
    &Ax = b 
    \\ &l\leq x \leq u
    \\ &x\in \mathbb{Z}^{n}
\end{align}
There exists a finite set $\mathbb{T} = \{t^{1},...,t^{N}$ that depends only on $A$ such that given a feasible, non-optimal point $x_{o}$, it satisfies for some $\alpha \in \mathbb{Z}_{+}$
\begin{enumerate}
    \item $f(x_{o} + \alpha t^{i}) < f(x_{o})$
    \item $x_{o} + \alpha t^{i}$ is feasible
\end{enumerate}
\subsection{Some Definitions}
\begin{itemize}
    \item The lattice \textbf{integer} kernel of $A$ is given by: \\ $\mathcal{L}(A) =  \left\{ x \vert Ax = 0, \: x\in \mathbb{Z}^{n}, A\in \mathbb{Z}^{m\times n} \right\}   \smallsetminus \{0\}$
    \item Partial ordering: $\forall x,y\in \mathbb{R}^{n} \: x\sqsubseteq y \: \textrm{iff} \: x_{i}y_{i}\geq 0 \: \& \vert x_{i}\vert \leq \vert y_{i} \vert \: \forall \: i = 1,2,...n $. If this condition is satisfied, x is said to be conformal to y. 
\end{itemize}
The Graver Basis is a very general and widely used test set. The Graver basis of $A$ is defined by:
\begin{align}
    \mathcal{G}(A) \subset \mathcal{L}(A) \subset \mathbb{Z}^{n} 
\end{align}
 such that $\mathcal{G}(A)$ is \textbf{$\sqsubseteq$ - minimal} ie, $\nexists x,y \in \mathcal{G}(A) \: s.t. \: x\sqsubseteq y$.
 \subsection{Applicability of the Graver Basis}
 The Graver Basis can  be used to augment feasible solutions and reach the optimum for the following objective functions $f(x)$:
 \begin{itemize}
     \item Separable convex minimization: $\sum_{i}f_{i}(c_{i}^{T}x)$ with $f_{i}$ convex
     \item Convex integer maximization: $-f(Wx)$ where $W\in\mathbb{Z}^{d\times n}$ and $f()$ is convex
     \item Norm p minimization: $f(x) = \vert \vert x - \hat x \vert \vert_{p}$
     \item Quadratic minimization: $f(x) = x^{T}Qx$ where $Q$ lies on the dual of the quadratic Graver cone of $A$ (This includes certain non-convex $Q\nsucceq 0$
     \item Polynomial minimization: $f(x) = P(x)$ where $P$ is a polynomial of degree $d$ that lies on cone $\mathbb{K}_{d}(A)$, the dual of the $d^{th}$ degree Graver cone of $A$
 \end{itemize}
 \subsection{Heuristics: the power of "Multi-seed"}
Mathematically, we need the complete Graver basis, a single feasible solution and an objective function of the form mentioned above to reach optimum points. However, Ising solvers might yield only a partial Graver basis and we also apply our algorithm on more complex objective functions, other than those described in the previous section. How does the Algorithm then work? Answer: Having \textbf{ multiple starting points} or "multi"-seeding. Its advantages:
\begin{itemize}
    \item We might obtain only a \textbf{partial Graver basis}. Since we have a large number of starting points, this increases the likelihood that at least one of them reaches the optimum.
    \item The objective function might be complicated. However, it can be approximated by our wide classes of functions given above. With a large number of seeds, we may cover various parts of the feasible space, and the odds of not reaching the minimum are quite low. 
\end{itemize}
\section{Quantum Classical Method to find the Graver Basis}
Find a binary (or unary) encoding for the variables to convert form an IP to a QUBO
\begin{align}
&\textrm{The integer program: }\\
&A x=0 \quad x \in \mathbb{Z}^{n} \quad A \in \mathbb{Z}^{m \times n}\\
&\min \quad x^{T} Q_{I}x, \quad Q_{I}=A^{T}A, \quad x \in \mathbb{Z}^{n}\\ \\
&\textrm{The Encoding: }\\
&x^{T}=\left[x_{1} \; x_{2} \cdots x_{n}\right] \; \quad
x_{i} \in \mathbb{Z}\\
&x_{i} = e_{i}^{T}X_{i}\\
&x_{i}^{T}=\left[X_{i, 1}\; X_{i, 2} \cdots X_{i, k_{i}}\right] \in\{0,1\}^{k_{i}}\\
&e_{i}^{T}=\left[2^{0}\; 2^{1} \ldots 2^{k i}\right]
\end{align}

Set a lower bound to allow for negative kernel values and reformulate:\\
\begin{align}
x=L+EX=\left[\begin{array}{c}
L_{x_{1}} \\
L_{x_{2}} \\
\vdots \\
L_{x_{n}}
\end{array}\right]+\left[\begin{array}{cccc}
e_{1}^{T} & 0^{T} & \ldots & 0 \\
0 &  e^T_{2} & \ldots & 0 \\
\vdots & \vdots & \ddots & \vdots \\
0 & 0 & \cdots & e_{n}^{T}
\end{array}\right]\left[\begin{array}{c}
X_{1} \\
X_{2} \\
\vdots \\
X_{0}
\end{array}\right]
\end{align}
Reformulate the original QUBO using this: \\
\begin{align}
    \min \quad X^{T} Q_{B} X, Q_{B} &=E^{T} Q_{I} E+\operatorname{diag}\left(2 L^{T} Q_{1} E\right) \\
    & X\in \{0,1\}^{nk}, \quad Q_{I} = A^{T}A
\end{align}
After reformulation, solve it with an Ising solver and invert previous transformations to obtain the kernel of $A$. After this, $\sqsubseteq$-minimal filtering done classically will yield a (partial) Graver Basis. 
\subsection{Finding the Graver Basis in 10 steps - A Toy Example}
This example will walk us through the abstract procedure that was described above.
\subsubsection{Matrix to Quadratic Unconstrained Integer Optimization (QUIO)}
Consider a Constraint Matrix $A$ given by: 
\begin{align}
A & = \left[\begin{array}{lll}
1 & 2 & 1
\end{array}\right] \\ 
A\,x & = 0 
\end{align}

We optimize the quadratic version of this:
\begin{align}
&\min \quad x^{T}Q_{I}x\\ 
&Q_{I}=A^{T}A=\left[\begin{array}{lll}
1 & 2 & 1 \\
2 & 4 & 2 \\
1 & 2 & 1
\end{array}\right] 
\end{align}
\subsubsection{Encode}
We can use a two bit encoding as shown:
\begin{align}
    e = \left[ \begin{array}{ll}
    2^{0} & 2^{1}
    \end{array}\right]
\end{align}
We will also shift one step to the left $L = -1$ to cover negative values.
\subsubsection{Obtain Encoding Matrix}
With all this in place, the encoding matrix is \\
\begin{align}
E = \left[\begin{array}{llllll}
1 & 2 & 0 & 0 & 0 & 0 \\
0 & 0 & 1 & 2 & 0 & 0\\
0 & 0 & 0 & 0 & 1 & 2
\end{array}\right]
\end{align}
\subsubsection{Obtain Encoded Equation}
Now, we have the transformation from $x$ to $X$ given by:

\begin{align}
x &= L + EX\\
\left[\begin{array}{l}
x_{1} \\
x_{2} \\
x_{3}
\end{array}\right]=\left[\begin{array}{c}
-1 \\
-1 \\
-1
\end{array}\right]&+\left[\begin{array}{llllll}
1 & 2 & 0 & 0 & 0 & 0 \\
0 & 0 & 1 & 2 & 0 & 0\\
0 & 0 & 0 & 0 & 1 & 2
\end{array}\right]\left[\begin{array}{l}
X_{1} \\
X_{2} \\
X_{3} \\
X_{4} \\
X_{5} \\
X_{6}
\end{array}\right]
\end{align}
\subsubsection{Formulate QUBO}
The QUBO can be formulated as follows:
\begin{align}
    & \min \quad (L+EX)^{T}Q_{I}(L+EX) \\ \\
    & \min \quad X^{T}\left (E^{T}Q_{I}E + 2\operatorname{diag}(L^{T}Q_{I}E)\right )X
\end{align}
Which boils down to:\\
\begin{align}
\min \left[\begin{array}{l}
x_{1} \\
x_{2} \\
x_{3} \\
x_{4} \\
x_{5} \\
x_{6}
\end{array}\right]^{\top}\left[\begin{array}{cccccc}
-7 & 2 & 2 & 4 & 1 & 2 \\
2 & -12 & 4 & 8 & 2 & 4 \\
2 & 4 & -12 & 8 & 2 & 4 \\
4 & 8 & 8 & -6 & 4 & 8 \\
1 & 2 & 2 & 4 & -7 & 2 \\
2 & 4 & 4 & 8 & 2 & -12
\end{array}\right]\left[\begin{array}{c}
X_{1} \\
X_{2} \\
X_{3} \\
X_{4} \\
X_{5} \\
X_{6}
\end{array}\right]
\end{align}

\subsubsection{Map to Ising Variables}
In this simple transformation, $S = 2X -1$, the Ising variable is $S$.
\subsubsection{Reformulate as Ising problem}
By substituting the above in $X^{T}QX$, we get a minimization of the form: \\
\begin{equation}
    min \quad S^{T}JS + h^{T}S
\end{equation}
\subsubsection{Solve Ising model and convert back to QUBO variables}
Now, we call our preferred Ising solver. In this lecture, we shall treat this as a black box which gives us a large number of optimal and near optimal samples for the given Ising Hamiltonian. We want as many zeros as we can get, so that we can extract more Graver elements. Hence, we can exploit the annealer (the sampler) since it can sample widely from the solutions. \\ The DWave quantum annealer gives some optimal and sub-optimal solutions. Of the latter, around 90\% of them have small overall sum-errors. We can postprocess these by adding and subtracting different sub-optimal vectors to get zero to maximize the number of Graver elements we procure. D-Wave performs really well for embeddable binary matrices and for narrow truncated band variables. \\ From this, we can convert back to QUBO variables from Ising variables ($X = (S+1)/2$, to get:
\begin{align}
[x] =\left[\begin{array}{lllllll}
0 & 0 & 1 & 1 & 0 & 0 & 1 \\
0 & 0 & 0 & 0 & 1 & 1 & 1 \\
0 & 0 & 0 & 1 & 0 & 1 & 0 \\
0 & 1 & 0 & 0 & 0 & 0 & 0 \\
0 & 0 & 1 & 1 & 0 & 0 & 1 \\
1 & 0 & 1 & 0 & 1 & 0 & 0
\end{array}\right]
\end{align}

\subsubsection{Decode and recover Kernel}
Invert the previously made transformations and recover $x$ as a vector of integers. 
\begin{align}
    x = L + EX =\left(\begin{array}{ccccccc}
-1 & -1 & 0 & 0 & 1 & 1 & 2 \\
0 & 1 & -1 & 0 & -1 & 0 & -1 \\
1 & -1 & 2 & 0 & 1 & -1 & 0
\end{array}\right)
\end{align}

\subsubsection{Convert Kernel to Graver Basis}
Now we do $\sqsubseteq$-minimal classical filteration to get $\mathcal{G}(A)$
\begin{align}
    \mathcal{G}(A)=\left(\begin{array}{cccc}
0 & 1 & 1 & 2 \\
-1 & -1 & 0 & -1 \\
2 & 1 & -1 & 0
\end{array}\right)
\end{align}
Negative basis elements are also part of the Graver basis, due to the symmetry in the constraint. So we can add $-\mathcal{G}(A)$ to our collection of Graver elements.   

\section{Finding feasible solutions and Augmenting}
Along the same lines, we can find several feasible solutions by similarly reformulating the constraints as shown below: 
\begin{align}
&A x=b  \quad l \leq x \leq u \\
&\min X^{T} Q_{B} X, \\  &Q_{B}=E^{T} Q_{I} E +2 \operatorname{diag}\left( L^{T} Q_{I}-b^{T} A\right) E \\
&X  \in\{0,1\}^{n k}, Q_{I}=A^{T} A
\end{align}
Solving this QUBO using an Ising solver will give us many feasible solutions and starting points. 
Adaptive centering and adaptive encoding width can be used for the feasibility bound, which makes the process faster. Once these points are found, we can augment each point in parallel using our (Partial) Graver Basis. These augmentations will lead us to the optimum point with high likelihood. 
\section{Demonstration of a Capital Budgeting problem}
In this \href{https://colab.research.google.com/github/bernalde/QuIP/blob/master/notebooks/Notebook\%206\%20-\%20GAMA.ipynb}{notebook}, the problem shown below was solved using GAMA:
\begin{align}
&\min -\sum_{i=1}^{n} \mu_{i} x_{i}+\sqrt{\frac{1-\varepsilon}{\varepsilon} \sum_{i=1}^{n} \sigma_{i}^{2} x_{i}^{2}}\\
&Ax=b \quad x \in\{0, 1\}^{n}
\end{align}
Key insights from the demonstration:
\begin{itemize}
    \item If we use a very small subset of the Graver Basis, we don't reach the optimum
    \item On the other hand, using anywhere between 10-100\% of the Graver Basis yielded very similar results over the entire range, provided we start at multiple feasible solutions
    \item Augmenting with the entire Graver Basis is time consuming and augmenting with a small subset leads to erroneous solutions. An appropriate trade-off has to be made. However, the entire Graver basis need not be used either.
    \item There are various ways of augmenting solutions like Greedy, Bisection methods, etc. They can be applied based on the situation at hand. 
\end{itemize}
Please refer to the notebook for more details about the implementation.
\section{Surpassing Best in Class Classical Methods}
Three areas where the D-Wave quantum computer is lacking are the number of qubits, their connectivity with each other and the coupling precision. 
\begin{itemize}
    \item If the precision is increased, we can run QUBOs over a larger range of parameters
    \item If the number of qubits increases, we can demonstrate GAMA on problems in which we know the state-of-art methods will not perform on par because of timing analysis for smaller problems. 
    \item GAMA can also surpass classical methods in problems with complex convex and non-convex objective functions.
\end{itemize}
\section{Classical Implementation of GAMA}
Suppose the matrix $A$ has a special structure. In that case, we can construct the Graver Basis from first principles and randomly generate many feasible solutions. This works very fast as compared to solvers like Gurobi. Problems belonging to such classes include QAP, QSAP, and CBQP. This work~ \cite{GAMA} describes this in more detail.

\section{Summary and Conclusion}
\begin{itemize}
    \item In conventional methods of solving Integer Programs, we quadratize the objective and add constraints to the objective with multipliers and optimize this entire mix.
    \item GAMA is a novel method where the objective function is used just as an oracle, and augmentation sets are found using annealers.
    \item It is a Quantum-Classical hybrid algorithm consisting of two parts, finding the Graver Basis and augmenting feasible solutions.
    \item The Graver Basis is created by solving for the Kernel using an annealer and classical post-processing to create a (partial) Graver Basis set.
    \item Many initial feasible solutions are found using an annealer and are later augmented classically
    \item Since we have many starting points, it is likely that the partial Graver Basis algorithm will find optimum values, even with complicated objective functions.
    \item There are some special cases where the Graver Bases can be computed classically, resulting in efficient performance compared to commercial solvers. 
    \item GAMA can thus be used in areas like Portfolio management, Cancer Genome Pathways, etc. \cite{Alghassi845719}, where we have a complex objective function subject to linear constraints.
\end{itemize}
\lecture{8}{Quantum Annealing}{Prof. Sridhar Tayur, David E. Bernal}{Shashank Kumar Ranu}
{https://bernalde.github.io/QuIP/slides/47-779\%20Lecture\%207\%20-\%20Quantum\%20Annealing.pdf}
{https://colab.research.google.com/github/bernalde/QuIP/blob/master/notebooks/Notebook\%207\%20-\%20DWave.ipynb}
{https://cmu.zoom.us/rec/share/9VBXMrjJZMt9U15BaJb49vXOGO2sV4HaHouNY-QsQmQWPoCLEmJ3BYjwbYR_ovmI.le7a5DFaS1nauKbq}
{2O8wq@Y\%}

\section{Qubit}
A qubit (or quantum bit) is the quantum-mechanical analog of a classical bit. In classical computing, information is encoded in bits. A bit can have one of two values zero or one. In quantum computing, information is encoded in qubits. A qubit is a two-level quantum system.  Consider a system
with 2 basis states, $\ket{0}$ and $\ket{1}$. The state of a qubit can be any superposition of these two basis states
$$\alpha\ket{0}+\beta\ket{1},\;\vert{\alpha}\vert^2+\vert{\beta}\vert^2.$$

\begin{figure}[h!]
\centerline{\includegraphics[width=0.35\textwidth,height=\textheight,keepaspectratio]{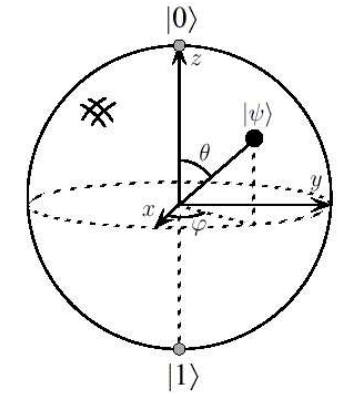}}
				\caption{Geometrical representation of Bloch sphere.}
				\label{fig:bloch} 
			\end{figure}

Using the constraints on $\alpha$ and $\beta$, a single qubit state can also be written as
$$\text{cos}\frac{\theta}{2}\ket{0}+e^{i\phi}\text{sin}\frac{\theta}{2}\ket{1}.$$
The numbers $0\leq \theta \leq \pi$ and
$0 \leq \phi \leq 2\pi$ define a point on a unit three-dimensional sphere. This is called
a Bloch sphere. The Bloch sphere is a geometric representation of qubit states as points on the surface of a unit sphere (see Fig. `\ref{fig:bloch}). Operations on single qubits that are commonly used in quantum information processing can be neatly described within the Bloch sphere picture.
\section{Superconducting qubits}
The ability to control individual quantum degrees of freedom and their interactions allows us to perform quantum computation. Different technologies such as
trapped atoms (ions or neutral), photons, single electrons trapped in silicon heterostructures, and magnetic/electric moments of molecules have been used to realise qubits.

Superconducting qubits are a leading approaches for realizing quantum logic elements and quantum coherent interactions with sufficiently high controllability
and low noise. Their design can be understood from the energy levels of an oscillator. The quantum harmonic oscillator shown in Fig.~\ref{fig:oscillator} is a resonant circuit comprising
of a capacitor and an inductor. For sufficiently low
temperature and dissipation, the
resulting harmonic potential supports quantized energy levels. However, due to the equidistant level spacing, the quantum harmonic oscillator by itself cannot be operated as a qubit. To remedy this situation, the circuit potential is made anharmonic by introducing a
nonlinear inductor – the Josephson junction. The imparted anharmonicity leads to a non-equidistant spacing of the energy levels, enabling one to uniquely address each transition, see Fig.~\ref{fig:oscillator}. Typically, the two lowest levels are used to define a qubit, with $\ket{0}$ corresponding to the ground state and $\ket{1}$ corresponding to the excited state. Large anharmonicity is generally favorable to suppress unwanted excitations to higher levels. 
\begin{figure}[h!]
\centerline{\includegraphics[width=0.3\textwidth,height=\textheight,keepaspectratio]{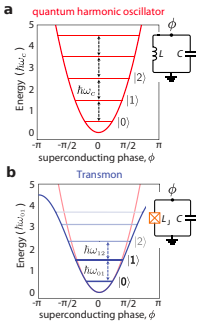}}
				\caption{Energy spectrum of a quantum harmonic oscillator.}
				\label{fig:oscillator} 
			\end{figure}
Different types of superconducting qubits exist in literature. D-wave annealers use flux superconducting qubits, while Google and IBM use transmon qubits for designing their quantum processors. The transmon qubit is a capacitively shunted variant of the Cooper pair box
that is largely insensitive to charge, resulting in improved reproducibility and coherence times. A flux qubit is based on magnetic flux whose states correspond to clockwise and counter-clockwise currents flowing around a loop interrupted by Josephson junctions. 
\begin{figure}[h!]
\centerline{\includegraphics[width=0.3\textwidth,height=\textheight,keepaspectratio]{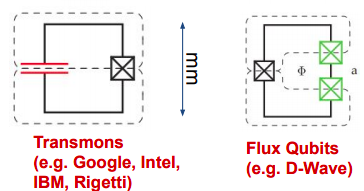}}
				\caption{Tansmon and Flux qubits architecture.}
				\label{fig:oscillator} 
			\end{figure}
.\section{The Quantum Adiabatic Algorithm (Quantum Annealing)}
Time evolution of a quantum system is fairly straightforward for a time-independent Hamiltonian. As per Schrodinger equation, the quantum state evolves as
\begin{equation}
    \ket{\psi(t)}=e^{-iHt}\ket{\psi(0)},
\end{equation}
where $H$ is the system Hamiltonian and $\ket{\psi(0)}$ is the initial state of the system.
Hence, an eigenstate $\ket{E}$ of the Hamiltonian acquires a time-dependent overall phase. 

Time evolution of a system having a time-dependent Hamiltonian is complex. However, if a Hamiltonian changes slowly, the dynamics remain simple and can be understood using the quantum adiabatic theorem. This theorem states that a quantum system which begins in a non-degenerate ground state of a time-dependent Hamiltonian will remain in the instantaneous ground state provided the Hamiltonian changes sufficiently slowly.
\section{The Quantum Adiabatic Algorithm for Ising Machines}
The quantum adiabatic theorem can be used to solve optimization problems. The first step is to map the objective function of the problem to the Hamiltonian of the quantum Ising system, given as
\begin{equation}
    H_P = \sum_{J_{ij}} Z_i\otimes Z_j+\sum_i h_i Z_i,
\end{equation}
where $J_{ij}$ captures the nearest neighbour interaction strength, and $Z$ is the Pauli Z matrix.

To apply the quantum adiabatic theorem: start from the minimum energy state of a problem with a known solution. One such Hamiltonian is the transverse field Hamiltonian 
\begin{equation}
    H_D=\tau \sum_i X_i.
\end{equation}
When acting upon $n$ qubits, this Hamiltonian has an equal superposition of all the $2^n$ possible binary strings as its minimum eigenstate. In D-wave annealers, analog signals control the implementation of these transverse and Ising models.  We start with a transverse field acting on all the qubits and quench it slowly to $0$. In parallel, we increase the flux signal controlling the Ising model Hamiltonian. At some point in time, both the signals co-exist, resulting in a complex dynamics. The analog signals are changed at a rate which depends upon the optimization problem being implemented. The quantum adiabatic theorem does not hold for the D-wave quantum annealer due to the system’s inherent noise. For example, mapping an objective function to the Ising Hamiltonian is not perfect due to the intrinsic control errors. Benchmarking and software mitigation of these errors is important to gain a proper understanding of the annealer output. However, D-wave annealer is still an adiabatic inspired quantum computer. 
\section{Minor Embedding}
Limited connectivity between the qubits is a restriction to employing the D-Wave quantum annealer for real-world applications. Before solving an optimization problem, it is necessary to map a problem graph onto a subgraph of the hardware graph. This process is called minor embedding. The problem graph is defined as a graph in which the vertices and edges represent the logical variables and interactions between them. The hardware graph is defined as a graph for which the vertices and edges represent the qubits and interactions between them, respectively. For example, suppose we have a QUBO with all non-zero quadratic coefficients. Implementing such a QUBO needs a fully connected annealer. 
\begin{figure}[h!]
\centerline{\includegraphics[width=0.7\textwidth,height=\textheight,keepaspectratio]{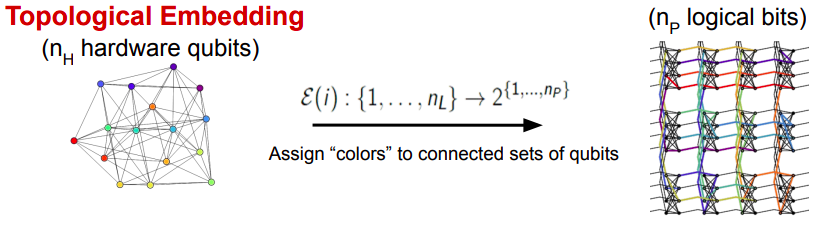}}
				\caption{An illustration of minor embedding.}
				\label{fig:embed} 
			\end{figure}
Minor embedding procedure consists of finding a set of connected subgraphs (logical qubits). These correspond
to different colors in Fig.\ref{fig:embe}, of the original graph such that each logical qubit can be associated with a node in the original graph. This association needs to be such that there is at least one edge between the qubits belonging to the associated logical qubits for every two connected nodes. However, such an embedding increases the qubit cost and changes the energy landscape. The problem of finding an optimal graph minor (i.e., minimizing the number of required nodes) is itself NP-hard and is typically tackled with heuristic approaches. For many graphs with a regular structure, an efficient embedding can be found systematically.
\begin{figure}[h!]
\centerline{\includegraphics[width=0.3\textwidth,height=\textheight,keepaspectratio]{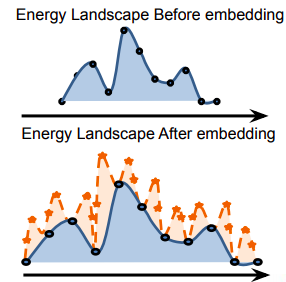}}
				\caption{Change in energy landscape due to minor embedding.}
				\label{fig:land} 
			\end{figure}

\section{Unembedding}
\begin{figure}[h!]
\centerline{\includegraphics[width=0.3\textwidth,height=\textheight,keepaspectratio]{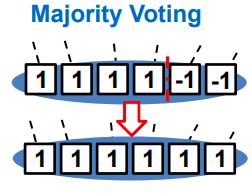}}
				\caption{Unembedding using majority voting.}
				\label{fig:majority} 
			\end{figure}

Modeling the connectivity structure of a given problem instance thus necessitates the computation of a minor embedding of the variables in the problem specification onto the logical qubits. One logical qubit consists of several physical qubits” chained” together. After annealing, it is, however, not guaranteed that all chained qubits get the same value ($-1$ or $+1$ for an Ising model, and $0$ or $1$ for a QUBO), and several approaches exist to assign a final value to each logical qubit, a process called unembedding. The most straightforward one is called majority vote. Suppose some logical qubit $x$
is mapped onto a chain with $m$ physical qubits. After annealing, the values of $m$ physical qubits are read, and $x$ is assigned the most common value among the $m$ chained qubits (see Fig.~\ref{fig:majority}). 

\section{Annealing Schedule Parameters}
\begin{figure}
\centerline{\includegraphics[width=0.4\textwidth,height=\textheight,keepaspectratio]{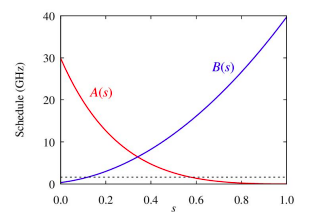}}
				\caption{Forward annealing.}
				\label{fig:anneal} 
			\end{figure}

Two energy scales are relevant during quantum annealing, $A(s)$ and $B(s)$. $A(s)$ represents the transverse, or tunneling, energy. $B(s)$ is the energy applied to the problem Hamiltonian. $s$ is the ratio between the current time during the anneal and the total annealing time and has a value between 0 and 1.
Energy scales $A(s)$ and $B(s)$ change during the quantum annealing process. In particular, a single, global, time-dependent bias controls the trajectory of $A$ and $B$. At any intermediate value of $s$, the ratio $A(s)/B(s)$ is fixed. We can choose the trajectory of one of $A$ or $B$ with time. The standard annealing schedule, $s$, produces a quadratic growth in $B(s)$. Typical values of $A(s)$ and $B(s)$ are shown in the Fig.\ref{fig:anneal}.
There is a {\em Reverse annealing} feature: this allows
one to start (in a classical state with the strength of the problem Hamiltonian and the driver Hamiltonians)  at the end of the  forward anneal, and to then increase the strength of the driver Hamiltonian and reduce that of the problem Hamiltonian, following the  schedule in reverse. One can pause, and then repeat this forward-backward approach gain.

\begin{figure}
    \centerline{\includegraphics[width=0.35\textwidth,height=\textheight,keepaspectratio]{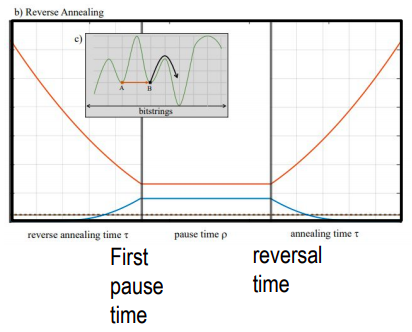}}
    \caption{Annealing with a pause.}
    \label{fig:pause} 
\end{figure}

The pause feature, which allows one to pause the anneal, keeping the strengths of the
driver Hamiltonian and the problem Hamiltonian constant for extended periods of time before completing the default annealing schedule (see Fig.~\ref{fig:pause}. A pause can increase the performance by orders of magnitude when the pause occurs within a well defined, relatively narrow region of the anneal but has little effect if placed outside that region.

\lecture{9}{Quantum Annealing and QAOA}{Prof. Sridhar Tayur, David E. Bernal, Dr. Davide Venturelli}{Shrigyan Brahmachari}
{https://bernalde.github.io/QuIP/slides/47-779\%20Lecture\%208\%20-\%20Quantum\%20Approximate\%20Optimization\%20Algorithm.pdf}
{}
{https://cmu.zoom.us/rec/share/B_AuEPTrdHhUerizLv_4QJ_2EcYlSr9oxqSbWso8ees7OOE8bWNnl3W5eeTVRoDS.Q7wNBj9JbEqRdapc}
{ghE58r=E}

\begin{itemize}
    \item minor embedding
    \item implementing quantum annealing and comparisons
    \item benchmarking study-QAOA
\end{itemize}

\section{Minor embedding}
Quantum annealing works by constructing a physical situation where the energy of the system maps to the objective function. Once the system to reaches a state of minimum energy, one obtains the minimum value of the objective function. 

A physical system's mathematical abstraction is a graph~( the adjacency matrix can be used to get the QUBO, as explained in the previous lectures).

However, the chimera graph~( subgraph of the chip) isn't necessarily complete~( that is, not all pairs of vertices are connected). In this case, we need to "duplicate" certain variables into several qubits/nodes~( representing decision variables)
This duplication step is non-trivial:
we can either use a heuristic method or solve a highly constrained problem.
Here is an example to illustrate this:
\begin{align}
z_i \in {1,-1}, \\
\min  ( z_1z_2+z_2z_3+z_3z_1)    
\end{align}
needs to be mapped into a subgraph which~( at least locally) has a square lattice-like structure.
Here is an equivalent mapping using constraints.:
\begin{align}
min (x_1x_2+x_2x_3+x_3x_4)\\
\text{such that} (x_1=x_4)    
\end{align}

However constraints can't be directly implemented. We need a QUBO form, so we use penalty functions:
\begin{align}
\min (x_1x_2+x_2x_3+x_3x_4+p(x_1-x_4)^2),\; \text{or}\\
\min (x_1x_2+x_2x_3+x_3x_4-p(x_1x_4)^2),
\end{align}
where the value of the chain strength $p>0$ should be large enough so the constraint is satisfied. However making $p$ very large will dilute the effect of other coefficients in the objective function.

\clearpage
\begin{figure}[tbh]
\centering{\includegraphics{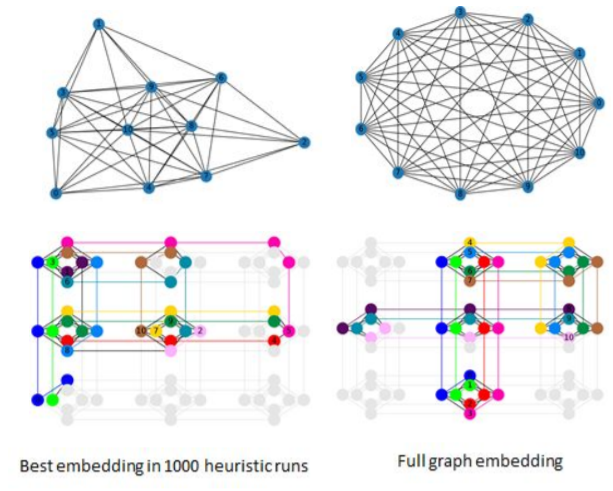}}
\end{figure}

\section{Implementing quantum annealing and comparisons}
Using classical Simulated Annealing with the default parameters increases the probability of finding the optimal solution from $(.5^{11})$ to 0.25 and a feasible solution to almost 1. The "infeasible" solutions are heavily penalized, and random sampling is not an option.

\begin{figure}[tbh]
\centering{\includegraphics{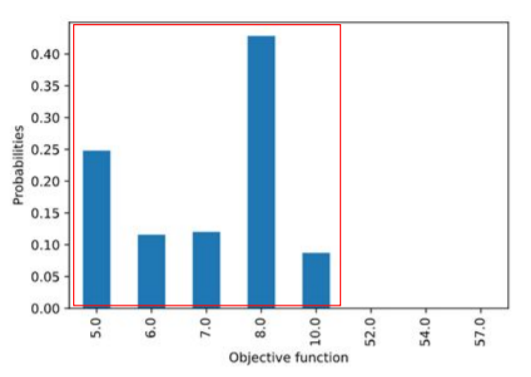}}
\end{figure}

Using Quantum Annealing with default parameters (annealing time, chain strength)
results in the probability of finding the optimal solution of 5/10000 and feasible of 15/10000.

\begin{figure}[tbh]
\centering{\includegraphics{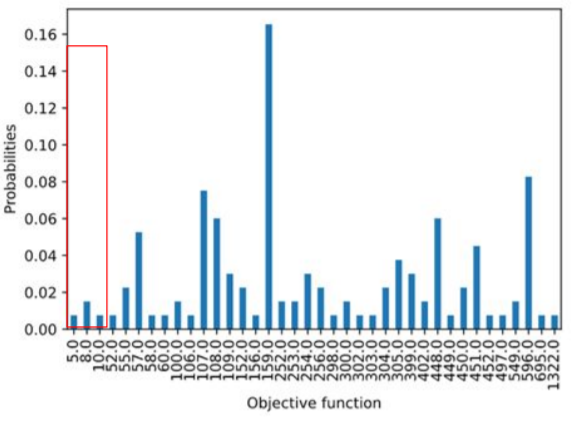}}
\end{figure}

In Quantum Annealing, we analyzed two different factors, the chain strength and the
annealing time. Our primary concern is maximizing the probability of success (feasible and optimal).

From the plots below, it is observed that there is a massive fraction of chain-breaking for small values of chain strength, i.e., the constraints aren't satisfied.
However, when we increase the chain strength ($p$), we see that the number of optimal solution initially increases and then falls. Here, $p$ can be decided by taking both plots into account.

\begin{figure}[tbh]
\includegraphics[width=0.5\columnwidth]{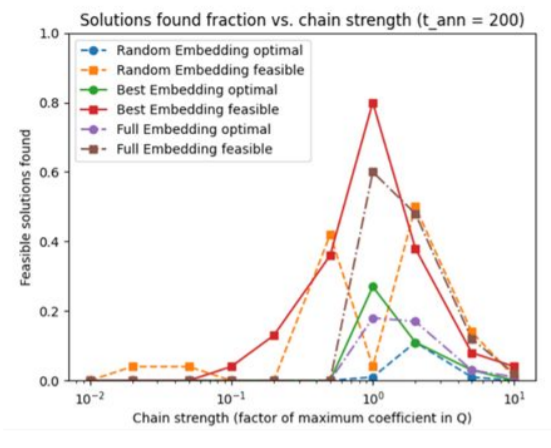}
\includegraphics[width=0.5\columnwidth]{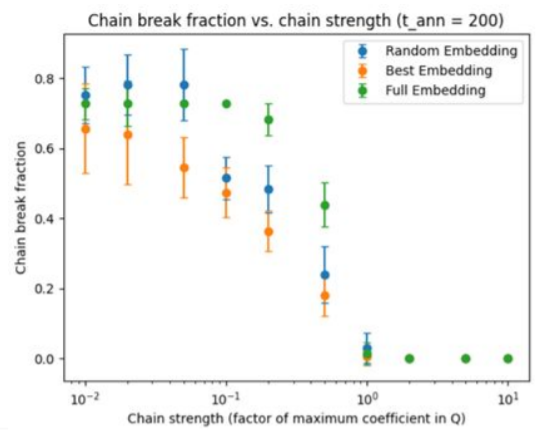}
\end{figure}

The following plots are for a shorter annealing time of 2~s, one-tenth of the previous set. While there is a drastic decrease in the number of optimal solutions~(the chain breaking seems unaffected), one should consider whether the increased running time is a better strategy than to run the annealing several times, somehow making sure different sets of solutions are found.

\begin{figure}[tbh]
\includegraphics[width=0.5\columnwidth]{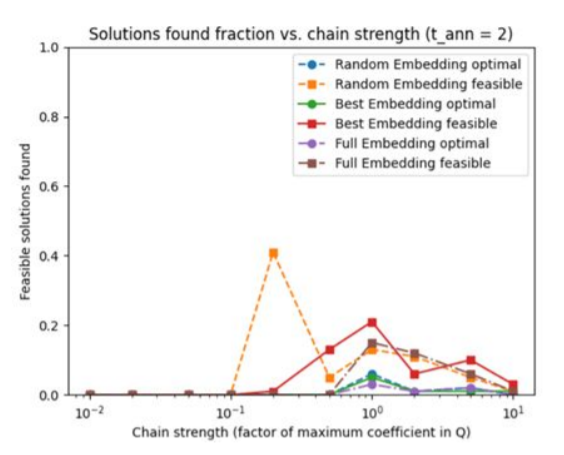}
\includegraphics[width=0.5\columnwidth]{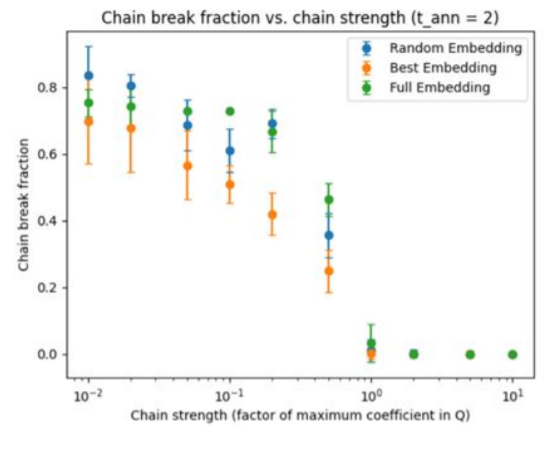}
\end{figure}

\section{How to do a benchmarking strategy}
There are 3 kinds of quantum speedup that are considered:
\begin{itemize}
  \item Provable Quantum Speedup (e.g. Grover)
    \item Strong Quantum Speedup (e.g. Shor)
    \item Quantum Speedup (potential, limited)
\end{itemize}
Note: it is possible to sometimes observe an apparent quantum speedup only because we may be dealing with small sets of test cases and the algorithms perform suboptimally. This does not give us the true asymptotic behavior.

Once we decide how to run, the benchmarking question is: What is the quality of a solution that I can expect for a random new instance with a given confidence?

\begin{itemize}
\item Set up the quantum algorithm on the QPU with some initial parameters
\item Run it several times and process the performance collecting the statistics of the distribution
\item If performance is not acceptable, use the distribution to choose new parameters~(might involve processing) 
$\rightarrow$ Repeat 1-3 times until satisfaction
\item Process final result and measure resources
used (time, energy)
$\rightarrow$ Repeat 1-4 for many benchmarking instances and collect the distribution of
performance.
\item Compare~(time, energy) against best classical method on available hardware 
\end{itemize} 

\begin{figure}[tbh]
\includegraphics[width=0.5\columnwidth]{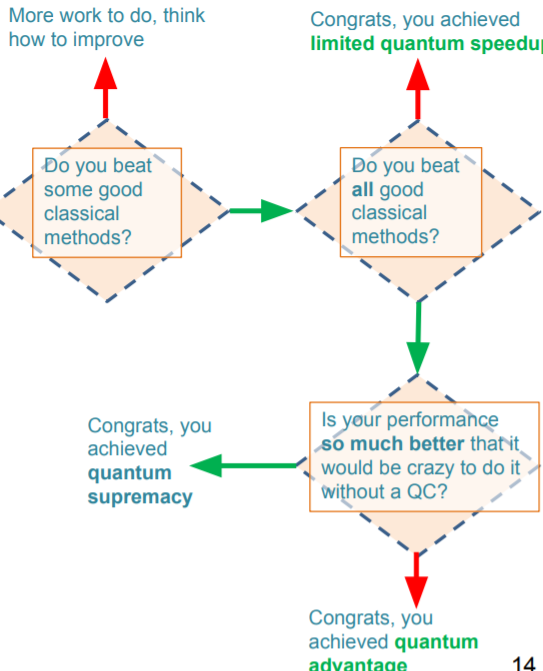}
\includegraphics[width=0.5\columnwidth]{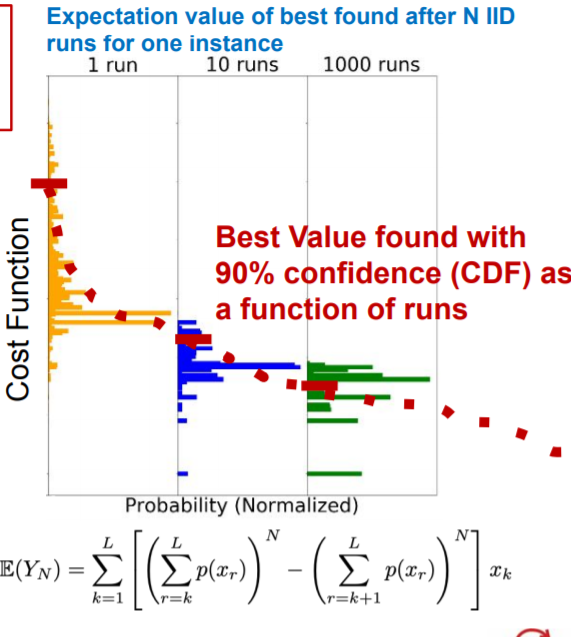}
\end{figure}

\section{Quantum Approximate Optimisation Algorithm}
\begin{itemize}
\item A quantum computer is UNIVERSAL if its instruction set allows the implementation of any algorithm allowed by quantum mechanics.
\item The time-evolution of the Ising model in a transverse field (Quantum Annealing as implemented in D-Wave) is NOT universal. However, the general AQC procedure is universal (need more complex $(H_p)$ and $(H_D)$). For quantum advantage in optimization
heuristics, universality is not necessarily required (the final state we are searching for is classical).
\end{itemize}
Why might you want a Universal Quantum Computer?
\begin{itemize}
\item Simulation of Quantum Systems
\item Flexibility in implementation of multiple quantum algorithms (e.g., Grover/Shor)
\item Exploit all the power of quantum mechanics
\item Making sure that what you do is not classically simulatable efficiently
\end{itemize}

QAOA operates using the concept of quantum circuits. Essentially, it can be shown that any unitary operation can be physically done to a qubit~(or a set of entangled qubits) because unitary operators can be shown to be an exponential of the time integral of a Hermitian, i.e., Hamiltonian operator. Using operators that work on several subsystems~(qubits), we can start with unentangled states into some desirable entangled states. To understand the kind of circuits and therefore operators one needs to get the desirable quantum entangled state, one uses linear algebra to decompose the transformation. Such decompositions are studied in detail in introductory course on Quantum Computing. It can be shown that any unitary transformation can be done by merely repeatedly using a recognized set of gates~(X, Hadamard, CNOT) in an appropriate sequence~(see Neilsen and Chuang).

QAOA aims to implement adiabatic transitions coherent operations more flexibly than AQC (digitally).  For infinite circuit this is at least as powerful as AQC.
 For finite circuit its power is unknown in general.  The following is an illustration of a simple example:
 \begin{figure}[tbh]
\includegraphics[width=0.5\columnwidth]{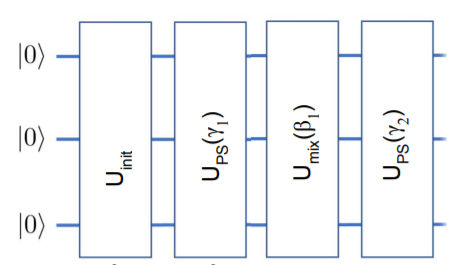}
\includegraphics[width=0.5\columnwidth]{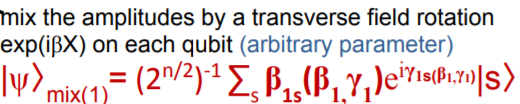}
\end{figure}

We start with some unentangled qubits; after the circuit transformation, the entangled state is such that when you measure the probability of a bit string, it
depends both on $\gamma$ and $\beta$ in a non-linear way. 

The primary difference to AQC is that the state that maps to the minimum energy is obtained as part of a superposition with other states. So there is a certain probability of realizing it by observation. Calculating this probability becomes very difficult.

\renewcommand{\thepage}{R-\arabic{page}}
\setcounter{page}{1}
\markboth{}{}
\bibliography{references}
\end{document}